\documentclass[a4paper,fleqn,usenatbib, twocolumn]{mnras}
\usepackage{newtxtext,newtxmath}

\usepackage[T1]{fontenc}
\usepackage{ae,aecompl}

\usepackage{siunitx}
\usepackage{mathrsfs} 
\usepackage{graphicx}	
\usepackage{amsmath}	
\usepackage{amssymb}	

\newcommand{\ket}[1]{{|#1\rangle}}
\newcommand{\bra}[1]{{\langle #1|}}

\newcommand{\dix}[1]{\cdot 10^{#1}}

\newcommand{\dif}[1]{\mathrm{d}#1} 
\newcommand{\difdeux}[1]{\mathrm{d}^2 #1} 
\newcommand{\diftrois}[1]{\mathrm{d}^3#1} 
\newcommand{\difv}[1]{\mathrm{d}^{3}\vect{#1}} 
\newcommand{\deriv}[2]{\frac{\mathrm{d}#1}{\mathrm{d}#2}} 
\newcommand{\derivdeux}[2]{\frac{\mathrm{d}^2#1}{\mathrm{d}#2}} 
\newcommand{\pderiv}[2]{\frac{\partial #1}{\partial #2}} 
\newcommand{\norm}[1]{\ensuremath{\left\Vert #1 \right\Vert}} 
\newcommand{\abs}[1]{\ensuremath{\left\vert #1 \right\vert}} 
\newcommand{\vect}[1]{\ensuremath{\vec{#1}}} 

\newcommand{\epsilonmin}{\epsilon_{min}}

\newcommand{\kparmin}[2]{\left(\begin{matrix} 
#1 \\
#2
\end{matrix} \right)
}

\newcommand{\deuxpi}{{2\pi}}

\newcommand{\unquart}{{\frac{1}{4}}}

\newcommand{\epsmax}{\epsilon_\text{max}}
\newcommand{\epsmin}{\epsilon_\text{min}}
\newcommand{\gdo}[1]{\bigcirc\left(#1\right)}

\newcommand{\zmin}{z_\text{min}}

\newcommand{\vx}{\vect{x}}
\newcommand{\vy}{\vect{y}}
\newcommand{\vz}{\vect{z}}

\newcommand{\vectk}{\vect{k}}

\newcommand{\pmax}{p_{\mathrm{max}}}

\title[Photon-photon pairs above a pulsar polar cap]{Electron-positron pair production by gamma rays in an anisotropic flux of soft photons, and application to pulsar polar caps}
\author[G. Voisin et al.]{
Guillaume Voisin,$^{1}$\thanks{E-mail: guillaume.voisin@obspm.fr (GV)}
Fabrice Mottez,$^{1,2}$
Silvano Bonazzola$^{1}$
\\
$^{1}$LUTH, Observatoire de Paris, PSL Research University, 5 place Jules Janssen, 92190 Meudon, France\\
$^{2}$LUTH, CNRS, 5 place Jules Janssen, 92190 Meudon, France\\
}

\date{Accepted XXX. Received YYY; in original form ZZZ}

\pubyear{2017}

\begin{document}
\label{firstpage}
\pagerange{\pageref{firstpage}--\pageref{lastpage}}
\maketitle

\begin{abstract}
Electron-positron pair production by collision of photons is investigated in view of application to pulsar physics.
    We compute the absorption rate of individual gamma-ray photons by an arbitrary anisotropic distribution of softer photons, and the energy and angular spectrum of the outgoing leptons.
   We work analytically within the approximation that  $1 \gg mc^2/E > \epsilon/E$, with $E$ and $\epsilon$ the gamma-ray and soft-photon maximum energy and $mc^2$ the electron mass energy. We give results at leading order in these small parameters. For practical purposes, we provide expressions in the form of Laurent series which give correct reaction rates in the isotropic case within an average error of $\sim 7\%$.
    We apply this formalism to gamma rays flying downward or upward from a hot neutron star thermally radiating at a uniform temperature of $10^6$K. Other temperatures can be easily deduced using the relevant scaling laws. We find differences in absorption between these two extreme directions of almost two orders of magnitude, much larger than our error estimate. The magnetosphere appears completely opaque to downward gamma rays while there are up to $\sim 10\%$ chances of absorbing an upward gamma ray. We provide energy and angular spectra for both upward and downward gamma rays. Energy spectra show a typical double peak, with larger separation at larger gamma-ray energies. Angular spectra are very narrow, with an opening angle ranging from $10^{-3}$ to $10^{-7}$ radians with increasing gamma-ray energies. 
   
\end{abstract}




\section{Introduction}

Electron-positron pair creation by collision of two photons, also called  Breit-Wheeler process, is important in a series of astrophysical questions \citep{Ruffini_2010}. Among them is the filling of recycled pulsar magnetospheres with plasmas.

The cross-section of two-photon-pair creation has been derived in \cite{berestetskii_quantum_1982}. This is a function 
of the four-momentum of both electrons. 
In pulsar magnetospheres, there is generally a huge reservoir of low-energy photons and a small number of high-energy photons. In order to decrease computational cost compared to pairwise calculations, the cross-section is integrated over the distribution of the low-energy photons. The exact formula for the reaction rate on an isotropic soft-photon background was first derived in \cite{nikishov_1962} to estimate the absorption of gamma rays by the extragalactic background light. Numerical integration was needed to obtain practical results. In contexts such as active galactic nuclei or X-ray binaries, various formulations and approximations were developed. Approximated analytical  expressions were given in \cite{bonometto_possible_1971} and \cite{agaronyan_photoproduction_1983} in the case of an isotropic soft-photon background distribution and averaging over outgoing angles of the produced leptons. The expression of \cite{agaronyan_photoproduction_1983} also applies for a bi-isotropic photon distribution (both strong and weak photon distributions are isotropic) without angle averaging over leptons. In these papers, the authors provide the energy spectrum of the outgoing leptons. An exact  expressions in the case of bi-isotropic photon distribution is derived in \cite{boettcher_pair_1997}, as well as a comparison to the previous approximations that favors the approach in \cite{agaronyan_photoproduction_1983} for its better accuracy. 

The standard picture of a pulsar magnetosphere assumes that its inner part is filled with plasma and corotates with the neutron star with angular velocity $\Omega_*$. The primary plasma is made of matter lifted from the neutron-star surface 
by electric fields \cite{Goldreich_Julian_1969}. These particles have highly relativistic energies; their motion in the neutron-star magnetic field generates synchrotron and curvature gamma-ray photons. In addition to primary particles, \citet{Sturrock_1971} has shown that electron-positron pairs are created in or near the acceleration regions of the magnetosphere. This provides plasma capable of screening the electric field component parallel to the magnetic field. There are two processes of pair creation : two-photon process, and one-photon in the presence of a strong magnetic field. 
The one-photon process is the most efficient with young and standard pulsars, of which magnetic field is in the range $B \sim 10^{6}-10^{8}$ T \citep{Burns_1984}. The photon-photon pair-creation process can become more important with high-temperature polar caps, and when the magnetic field is below $10^{6}$ T as in recycled pulsar magnetospheres. Anisotropy of the soft-photon sources is prone to be important as they are expected to come either from the star (hot spots) or from synchrotron radiation in magnetospheric gaps. That is the main reason of our present investigation.

Many detailed studies of pair-creation cascades in pulsar magnetospheres are based only on the one photon process. This is for instance the case in the recent studies in \citet{Timokhin_2015_1}. Others take the two reactions into account \citep{Chen_Beloborodov_2014,harding_regimes_2002}.

In numerical simulations of pulsars, the pair-creation rate is generally estimated with simple proxies. For instance, in \cite{Chen_Beloborodov_2014}, a mean free path $l=0.2 R_*$ is used for the one-photon process, and  $l= 2 R_*$ for the two-photon process. The rate of creation of pairs is not explicited as a function of the electron (or positron) momentum, neither of the local photon background. Instead, pair creations are supposed to be abundant enough to supply electric charges and current densities. The authors write that this approximation is somehow similar to the force-free approximation. 
In \citet{harding_regimes_2002}, both one-photon and two-photon processes are taken into account, and the two-photon process is controlled by a mean free path derived from \cite{zhang_two-photon_1998}, where anisotropy is partially taken into account : the energy integral has a lower limit that depends on the angular size of the hot cap providing the soft-photon background.  Besides, these authors do not provide spectra for the created pairs although the energy distribution of the outgoing particles are important for the dynamics of pair cascades. A more complete model needs an integration over every local surface element with a threshold that depends on the location of each elementary emitter. This is what the results of the present paper allow to do within some approximation, together with angular and energy spectra of the outgoing pairs.

Pair creation by two photons is also important in high energy gamma-ray astrophysics. Many papers about gamma-ray bursts and active galactic nuclei refer to \cite{Svensson_1987} and the integrated mean free path in this paper is also based on \citet{nikishov_1962}. Actually, spectra of TeV radiation observed from distant (beyond 100 Mpc) extragalactic objects suffer essential
deformation during the passage through the intergalactic medium, caused by energy-dependent absorption of
gamma rays interacting with the diffuse extragalactic background light \citep{nikishov_1962,gould_opacity_1966}. 
This effect drastically limits the horizon of the gamma-ray universe, and this has been taken into account in the science case of high-energy gamma ray observatories \citep{Vassiliev_2000}. 

In this paper, we revisit the computation of the two-photon pair-creation rate with the aim of dealing with arbitrarily anisotropic soft-photon background distribution. In addition, we give formulas for angle and energy spectra in order to be able to determine in which state pairs are created. After an introduction to the two-photon pair creation equations in section \ref{sec_two_photon}, the integral over the low-energy photons is defined in section \ref{ssLmoins}. 
Practical expressions for spectra are derived in section \ref{sec:solution}, and applications to the cosmic microwave background and to a hot neutron star are developed in section \ref{sec:pulsar}. 


\section{The two-photon-quantum-electrodynamics reaction} \label{sec_two_photon}

When not specified, we use a unit system where the speed of light $c=1$.

\subsection{General formalism \label{sec_two_photon_general}}

Any quantum-electrodynamics reaction from an initial quantum state $\ket{i}$ to an outgoing state $\ket{o}$ can be represented as the
 decomposition on a final states basis $\{\ket{f_k}\}$ of the evolved state $\hat{S}\ket{i}$, $\hat{S}$ being the evolution operator,
\begin{equation}
	\label{eqpairpaircreation}
	\ket{o} = \sum_k \bra{f_k}\hat{S}\ket{i} \ket{f_k}
\end{equation}
From that starting point, if one is able to derive the appropriate evolution operator, one can then determine the probability of transition from a given state to any state of the final basis. 
We are interested in the reaction which yields an electron $e_-$ and a positron $e_+$ from the encounter of two photons. Common applications take place in a frame where one is "strong", that is high-energy, and the other is "weak". Hence we call them $\gamma_s$ and $\gamma_w$, and
\begin{equation}
\label{ggee}
 \gamma_s + \gamma_w \rightarrow e_- + e_+
\end{equation}

\begin{figure}
	\begin{center}
		\def\svgwidth{0.45\textwidth}
	    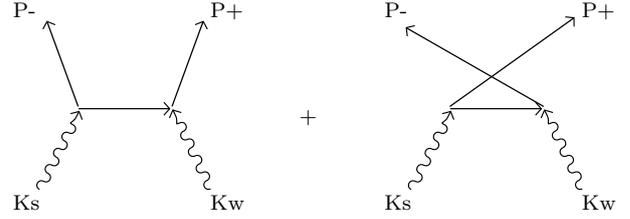
	 \end{center}
	 \caption{Reaction of electron-positron pair creation from a pair of photons represented to first order by Feynman diagrams. Photons have 4-momenta  $K_s$ and $K_w$ while electron and positron have respectively $P_-$ and $P_+$.}
\end{figure}

The state of a free photon can be decomposed on a plane-wave basis parametrized by four-momentum and polarisation. The common assumption is that the effective state of a photon is very well approximated by one plane wave at the time of the encounter. Such a state is not physical in itself, because it cannot be normalized i.e. it does not belong to the $L^2$ space, or more physically  because the Heinsenberg inequality imposes to the wave function to be entirely spread through space as a consequence of the perfect determination of momentum. Though, this assumption should be valid over a local four-volume of space-time $V\delta t$  where the interaction through operator $\hat{S}$ takes place.

Equivalently, free electrons and positrons live on a basis of plane-wave spinors parametrized by a four-momentum and a spin. 
From now on, the leptons are characterized by their charges and their four-momenta $P_+ = (P^0_+, \vect{p_+})$ and $P_-= (P^0_-, \vect{p_-})$  and photons by their four-momenta $K_s = (K^0_s, \vect{k_s})$, and $K_w= (K^0_w, \vect{k_w})$. We consider that their distributions are averaged over spin and polarization respectively. 
Following \cite{berestetskii_quantum_1982}, the Lorentz-invariant cross-section equations are derived in terms of kinematic invariants (also called Mandelstam variables), defined as
\begin{align}
	\label{eqKininvariants}
	s = (P_- - K_s)^{2} = (P_+ - K_w)^{2}, \\  \nonumber
	t = (K_w + K_s)^{2} =  (P_+ + P_-)^{2}, \\ \nonumber
	u = ((P_- - K_w)^{2} =  (P_+ - K_s)^{2}. 
\end{align}
The conservation of four-momentum writes
\begin{equation}
	s + t + u = 2m^2,
\end{equation}
where $m$ is the mass of the electron. 

The probability $\dif{w}$ per unit time of making a pair is
\begin{equation}
	\dif{w} = \dif^2{\sigma} \times j,
\end{equation}
where  $\dif{\sigma}$ is the Lorentz-invariant cross-section
\begin{eqnarray} \label{eq_cross_section_pair}
\dif^2{\sigma} &=& - \dif{s} 8\pi r_e^2 \frac{m^2}{t^2} \times  \left[ \left( \frac{m^2}{s-m^2} + \frac{m^2}{u-m^2}\right)^2 + \right.\\  \nonumber 
&&\left. \left( \frac{m^2}{s-m^2} + \frac{m^2}{u-m^2}\right) - \frac{1}{4}\left(\frac{s-m^2}{u-m^2} + \frac{u-m^2}{s-m^2}\right) \right], 
\end{eqnarray}
where $ r_e $  is the classical radius of the electron \footnote{In international units $r_e = \frac{e^2}{4\pi\epsilon_0 m c^2} \simeq 2.8179 \dix{-15} $ meters, with $\epsilon_0$ the electric permittivity of vacuum.},  $\dif{s}$ is the differential of the kinematic invariant $s$ at $P_-^0$ and $K_s^0$ fixed, 
\begin{eqnarray}
\dif{s}=2\dif{(\vect{p_-}\cdot\vect{k_s})} =  2\norm{\vect{p_-}} \norm{\vect{k_s}} \sin (\vect{p_-},\vect{k_s})  \dif{(\vect{p_-},\vect{k_s})}
\end{eqnarray}
and $j$ is the elementary two-particle flux of the reaction,
\begin{equation} \label{eqDefinitionJ}
	j = \frac{1}{V}\frac{K_s\cdot K_w}{K^0_s K^0_w},
\end{equation}
and $V$ is the interaction volume previously defined \footnote{include a factor $c$ in the definition of $j$ when it is not assumed that $c=1$.}. 
Only the current $j$ is frame-dependent. In particular it reads $ j = 2/V $ in the center of mass (CM) of the reaction.

Let us notice that a reaction is possible only if the energy of the two photons exceeds the mass energy of the electron and of the positron. One shows that the kinematic invariant $t$ \eqref{eqKininvariants} is equal to the square of the energy in the CM. This allows to define the frame-invariant criterion 
\begin{equation}
\sqrt{t} \geq 2m,
\end{equation}
which turns into 
\begin{equation} \label{eq_threshold}
K^0_s K^0_w\left(1-\cos\xi\right) \geq 2m^2
\end{equation}
where $\xi$ is the angle between the two photons. 

A few important properties of $\dif{w}$ can be evidenced by taking a look at cross-section\eqref{eq_cross_section_pair} averaged over every possible direction of the outgoing lepton \citep{berestetskii_quantum_1982}. As a result, the averaged cross-section depends only on the kinematic invariant $\tau= t/(4m^2)$ the ratio between the CM energy and the threshold energy. Without loss of generality in the present discussion, we can assume that the reaction takes place in the CM frame, such that the elementary current $j= 2/V$.  The ultra-relativistic limit \citep{berestetskii_quantum_1982} shows that the cross-section vanishes like $\log\tau/\tau$. This kind of decrease with energy is a common feature of quantum mechanical cross-sections. Moreover, one can numerically estimate the CM energy corresponding to the maximum of the reaction rate to be $\sqrt{t} \simeq 1.4(2m)$. 

Concerning the angular dependency, leptons are created almost isotropically when the reaction is near threshold while their momenta become aligned with those of the progenitor photons when going to higher energies (see e.g. \citet{berestetskii_quantum_1982}). In the observer's frame this translates in a larger angular dispersion for reactions close to threshold.

Equations (\ref{eqKininvariants}-\ref{eqDefinitionJ}) fully describe the interaction for a given pair of photons; but in a pulsar's magnetosphere, there is a huge amount of photon pairs. In a simulation, it is not possible to compute $dw$ for each pair; we need a statistical approach and a kind of "collective" reaction rate $dW$. We define it in the next section.

\subsection{The pair reactions that count in a pulsar magnetosphere}
In a pulsar magnetosphere, the weak photons $K_w$ are mostly caused by the black-body radiation of the neutron star, or possibly by synchrotron from secondary pair cascades. Their energies range in the X-ray domain. The strong photons are caused by the synchro-curvature radiation of energetic particles (electrons, positrons, and possibly ions). Their energies are in the gamma-ray domain. They are more scarce than weak photons.
Let's follow a "rare" high-energy, strong photon taken from a phase-space distribution $f_s$. We assume that it flies through an abundant stream of low-energy, weak photons with a distribution $f_w$. 
	Strong photons negligibly interact with other strong photons because they are not abundant, and because the reaction would likely be far above threshold in Eq. (\ref{eq_threshold}), and therefore inefficient.
	Weak photons do not interact with other weak photons since their energies are under the reaction threshold. 
	Thus, only weak/strong interactions remain, but weak photons are so numerous that a reaction negligibly changes their distribution.
	Because strong photons are less abundant, pair creations can change their distribution. 
Hence, for the simulation of a pulsar's magnetosphere, we need to compute the probability of interaction of a strong photon on the background distribution $f_w$ of weak photons. Indeed, it does not matter which weak photon is annihilated  but we want to update $f_s$ as well as the lepton distributions with the outcome of the reactions. With our representation of the involved particles, this amounts to compute the probability $\dif{W}$ of creating a lepton of four-momentum $P$ from a photon $K_s$,
\begin{equation}
	\dif{W} = \dif{W}_{K_s \rightarrow P}.
\end{equation}
For example, one could think of high-energy synchrotron or curvature photon emitted above the polar cap of a pulsar and flying through a stream of thermal photons emitted by the crust.
Let us notice that 
	the probabilities of making a positron or an electron are the same, 
	and that a four-momentum has four components but only three are independent since $\norm{P}^2 = m^2c^4$. These three free parameters can be parametrized by one direction (two parameters) and the energy of the particle. 

\section{Probability of reaction for a given photon distribution} \label{ssLmoins} 

Quite naturally, the desired probability is the sum over all the possible reactions involving a photon $K_s$ from the background, that would produce an electron at $P_-$ (respectively a positron at $P_+$),
\begin{equation}
	\dif{W_{f_w}(K_s, P_-)} = \sum_{L_- = \{(K_w, P_+) / K_s \rightarrow P_-\}} \dif{w}(K_w, K_s, P) \times  N_w  \times N_s,
\end{equation}
where $N_w$ and $N_s$  are the number of photons of four-momentums $K_w$ and $K_s$ respectively within the interaction volume $V$. 
In spite of greater simplicity in the CM frame, we must use Eq. \eqref{eq_cross_section_pair} in the laboratory frame, because the CM frame would be different for each of the summed pairs of photons. 
Since low energy photons are parametrized continuously, we must change our sum for an integral, which yields
\begin{eqnarray} \label{eq_integrales}
	N_w & \rightarrow & f_w(\vect{x_w}, \vect{k_w}) \difv{x}\difv{k_w}, \nonumber\\
	\dif{W_{f_w}(K_s, P_-)} & = & N_s \int_{L_-}   \frac{c}{V}\frac{K_s\cdot K_w}{K^0_s K^0_w} 
	 f_w(\vect{x}, \vect{k_w}) \dif^{6}{\Omega}, \nonumber \\
	\mbox{ where }\dif^{6}{\Omega}& = & \difdeux{\sigma} \difv{x}\difv{k_w}.
\end{eqnarray}


We assume that strong photons are spread out in space such that their density does not vary on the interaction volume $V$ such that their local density is $n_s = N_s / V$. Consequently the differential probability of interaction per unit time reads 

\begin{eqnarray}
\label{eqWoverV}
	\dif{W_{f_w}(K_s, P_-)}  &=&  n_s  W_k,  \\ \nonumber
	W_k&=&\int_{L_-} \difdeux{\sigma} \frac{c K_s\cdot K_w}{K^0_s K^0_w} f_w(\vect{x}, \vect{k_w})  \difv{k_w},
\end{eqnarray}
where the volume element $\dif{V} = \difv{x}$. 


\subsection{The domain of integration} 

Let us precisely define the domain $L_-$ of integration. 
We note $\Pi_\alpha = \{ P \in \mathbb{R}^4 : \norm{P} = \alpha^2c^4, P^0 \geq \abs{\alpha}\}$ such that $\Pi_m$ is the set of lepton four-momenta (m being the mass of the electron) and $\Pi_0$ is the set of photon four-momenta. Then,
\begin{equation}
	\label{naivelm}
	L_-(P_-,K_s) = \{ (P_+, K_w) \in \Pi_m\times\Pi_0 : K_w - P_+ = P_- - K_s \}.
\end{equation}
Equivalently, $L_-$ is the subset of $\mathbb{R}^4 \times \mathbb{R}^4$ parametrized by $K_w$ with the following constraints:
\begin{equation}
\label{eqConstraintsLm}
	\left\{
	\begin{array}{rclr}
		P_+ & = & K_w - (P_- - K_s) & \mathrm{(a)},\\
		\norm{P_+}^2 & = & m^2 c^4 & \mathrm{(b)}, \\
		\norm{K_w}^2 & = & 0   & \mathrm{(c)},  \\
		K_w^0 & \geq & 0 & \mathrm{(d)},\\
		P_+^0 & \geq & m c^2&		\mathrm{(e)}.
	\end{array} \right.
\end{equation}
Condition (a) expresses the conservation of four-momentum. Conditions (b) and (e) come from $P_+^0 \in \Pi_m$, and conditions (c) and (d) come from  $K_w \in \Pi_0$.
We can compute the number of degrees of freedom in $L_-$. The set $L_-$ is a subset of $\Pi_m \times \Pi_0$ of dimension 8.
The condition (a) on quadrivectors substracts 4 degrees of freedom. The conditions (b) and (c) both substract 1 degree of freedom. We are left with a set $L_-$ of dimension 2. 

Some of the conditions in Eq \eqref{eqConstraintsLm} are already incorporated in the solution of our problem.
Condition (c) is already implicitly met in Eq. (\ref{eqWoverV}). Condition (a) is also implicitly met by the set of variables used. Only (b) is not straightforward, since $P_+$ is not directly part of the variables of integration. One can still convert it into a condition on the three other four-vectors by putting (a) into (b) and using (c), the three following equalities being equivalent:
\begin{eqnarray}
	\label{eqmassshell2}
	\norm{P_+}^2  =  m^2 c^4 \nonumber  \\  \norm{K_w - (P_- - K_s)}^2 =  m^2c^4 \\
		  K_w^0(K_s^0 - P_-^0) - \norm{\vect{k_s} - \vec{p_-}} K_w^0 \cos\xi  = K_s\cdot P_-, \label{eqconstLc} \\
		  \xi = \mbox{angle}({\vect{k_s} - \vec{p_-},\vect{k_w}})
\end{eqnarray}
where  $\norm{\vect{k_w}} = K_w^0$.
The limit case where $\vect{k_s} = \vec{p_-}$, for which $\cos\xi$ is not defined, is physically impossible  because Eq. (\ref{eqconstLc}) would imply $K_w^0<0$, in contradiction with condition (d). 
%
With some algebra, we can show that $K_s\cdot P_- \ge 0$ and that the condition $\abs{\cos\xi} \leq 1$ imposes $K_w^0 > \epsilonmin$, where 
\begin{equation} \label{eq_epsilonmin}
\epsilonmin =  \frac{K_s\cdot P_-}{\norm{\vect{k_s} - \vect{p_-}} + K_s^0 - P_-^0}
\end{equation}
More precisely $K_w^0([-1,\cos\xi_0[) = [\epsilonmin,+\infty[$ and $K_w^0(\cos\xi > \cos\xi_0) < 0$.


We can distinguish three regimes of approximation:
\begin{eqnarray}
	K_s^0 >> p_- & \mathrm{ : } & \epsilonmin \sim \sqrt{m^2 + p_-^2} - p_-\cos\theta, \\
	K_s^0 \sim p_- & \mathrm{ : } & \epsilonmin \sim K_s^0\sqrt{\frac{1 - \cos\theta}{2}} \nonumber, \\ 
	K_s^0 << p_- & \mathrm{ : } & \epsilonmin \sim	 p_-.
\end{eqnarray}
For further approximations, we consider that the weak-photon distribution has a cut-off 
at $\epsilon = \epsmax < m/4$.
\begin{eqnarray}
\label{mainapprox}
	k \gg m/4 > \epsmax =128 \mbox{keV}.
\end{eqnarray}
Because the weak distribution function is in the most extreme case composed of thermal X-rays typically in the range $1-10$ keV for a pulsar, this approximation is reasonable.

\section{General solution\label{sec:solution}}
\subsection{Energy spectrum}
The probability of interaction depends on the integral $W_k$ defined in Eq. \eqref{eqWoverV}. 
In this section, $W_k$ is directly  expressed as a multiple integral with explicit boundaries. The results exposed in this section can be used directly for applications. 
The path followed to compute them are described in appendix \ref{sec_correct_rate}. A summary of the notations and useful relations is given in appendix \ref{apformulae}.
The new expression of $W_k$ involves new variables that appear both in the integrand and in the boundaries of the integral. 
First, new notations are introduced, for shorter formulas,
\begin{eqnarray}
 K_s & \equiv & \left(k, \vect{k} = k\vect{z} \right), \\
 P & \equiv & \left(P^0, \vect{p}\right), \\
 K_w & \equiv & \left(\epsilon, \vect{x} =(x,y,z) \right), \\
 \theta & \equiv & \mbox{angle}(\vect{k},\vect{p}).
\end{eqnarray}
With the new notations related to $P_-$ and to $K_s$, the integration set $L_-(P_-,K_s)$ can be rewritten
$L_-(p, \cos \theta,k)$.
Let $\Omega$ be the set of angular components of the 
electron $P_-$, we rewrite $W_k$ as
\begin{equation}
\label{eqWapprox1}
	W_{\vectk} = c \int_{\Omega} \dif{\Omega} \int_{L_-} \derivdeux{\sigma}{\Omega} \frac{K_s\cdot K_w}{K_s^0 K_w^0} f_w(\vect{k_w}) \diftrois{\vect{k_w}}.
\end{equation}
We wish to compute the probability of making a pair of which the electron $P_-$ is in a volume of phase space defined by
\begin{eqnarray} \nonumber
k/2 < p_1 < & p & < p_2 < k \\  \nonumber
(\cos\theta, \phi) \in \Omega & = & [C_1, C_2 ] \times [0,\deuxpi ] \text{ with } C_\text{min} \leq C_1 < C_2 < 1
\end{eqnarray}
where $p_1, p_2, C_1$ and $C_2$ can be set arbitrarily as long as the above inequalities are correct.
After the computations exposed in section \ref{sec_correct_rate}, $W_k$ is transformed into a multiple integral with explicit boundaries.
Before showing it, a new set of variables is introduced. 
The parameter $\mu$ parametrizes $\cos\theta$,
\begin{equation}
\label{eqmudef}
\cos\theta = 1 - \left( \frac{2(k-p)}{kp}\mu\epsmax - \frac{m^2}{2p^2} \right).
\end{equation}
\newcommand{\mumin}{\mu_\mathrm{min}}
It varies in an interval $\mu \in [\mumin, 1] $ where $\mu = 1 $ corresponds to $\cos\theta = c_\text{min}$ and $\mumin $ is such that $\cos\theta = 1$,
\begin{equation}
\label{eqmumin}
\mumin = \unquart\frac{km}{p(k-p)} \frac{m}{\epsmax}
\end{equation}
We define the dimensionless coefficients $a_i(p)$,
\begin{eqnarray} \label{eq_coefs_a_i}
	a_1(p) & = & -\frac{m^2 \left(k^2-2 k p+2 p^2\right)}{8 k p^2 (k-p)}, \\
	a_2(p) & = & -\frac{m^4 \left(k^2-4 k p+2 p^2\right)}{16 \epsmax k p^3 (p-k)}, \\ 
	a_3(p) & = & \frac{m^6 (3 k-2 p)}{32 \epsmax^2 p^3 (k-p)^2},\\
	a_4(p) & = & -\frac{k m^8}{64 \epsmax^3 p^4 (k-p)^2},
\end{eqnarray}
and
\begin{equation}
\label{eqR_texte_principal}
R = 2\epsmax\sqrt{\mu (1 - \mu)}.
\end{equation}
In the following we do not write the $p$ dependance of the $a_i$ coefficients except when otherwise stated. 
It is convenient to express the weak-photon three-momentum in cylindrical coordinates $\vect{x} =(r,\phi_w, z)$. Then, only the distribution function $f_w$ depends on the angle $\phi_w$, which allows a direct integration defining the function ( see also \eqref{eqaveragephiw})
\begin{equation}
	\label{eq:Fwdef}
	F_w(r, \mu) = \int_{\phi_w = 0}^\deuxpi f_w\left(r, \phi_w, z(r^2,\mu)\right) \dif{\phi_w},
\end{equation}
where $z(r^2,\mu)$ is defined in equation \eqref{eqzapprox} by 
\begin{equation}
\label{eq:zdefintext}
z(r^2,\mu) = \frac{k}{4}\left(\frac{r^2}{\mu k\epsmax} - 4 \mu\frac{\epsmax}{k}\right).
\end{equation}
The integral $W_k$ in \eqref{eqWapprox1} is approximated by
\begin{eqnarray}
	\label{eqWapproxsplitreduced}
	W_{\vectk} = c\deuxpi \int_{p_1}^{p_2} \dif{p} \int_{\mu_1}^{\mu_2} \dif{\mu} \sum_{i=1}^{4}  \frac{a_i}{\mu^i} \int_{r = 0}^{R} 2  F_w(r, \mu)r \dif{r}.
\end{eqnarray}
Here, the boundaries of the integration domain are left arbitrarily. 
The reaction probability integrated other every outgoing momenta can be computed as well.
In this case the $\mu$ integral is taken from $\mumin$  
to 1 and $p$ ranges between $k/2$ and  a maximum $\pmax$ 
defined such as  $\mumin(p = \pmax) = 1$. We find
\begin{equation}
\pmax = \frac{k}{2}\left(1 + \sqrt{1 - \frac{m^2}{k\epsmax}}\right) 
\end{equation}
The spectrum of outgoing lepton energy is readily obtained as 
\begin{equation} \label{eqWfoutgoingleptonenergy}
\deriv{W_{\vectk} }{p}\left(\max\left(p, k-p \right)\right) = c\deuxpi \int_{\mumin}^{1} \dif{\mu} \sum_{i=1}^{4}  \frac{a_i}{\mu^i} \int_{r = 0}^{R} 2  F_w(r, p, \mu)r \dif{r}.
\end{equation}

\subsection{Angular spectrum}

It is also possible to compute the angular spectrum of the outgoing leptons. The problem has to be split in two, whether one consider the higher-energy particle ($p > k/2$) or the lower-energy particle ($p < k/2$). 
\newcommand{\cthet}{{c_\theta}}
\newcommand{\pmin}{p_{\min}}

For the higher-energy particle, one takes equation \eqref{eqWapproxsplitreduced} and changes variable $\mu$ to $\cthet = 1-\cos\theta$ using equation \eqref{eqmudef}. One then obtains 
\begin{equation}
\label{eq:dwdc}
\deriv{W_{\vect{k}}}{\cthet} = \deuxpi c \int_{p_1}^{p_2} \dif{p}\sum_{i=1}^{4} \frac{a_i'}{c^i}\int_{r=0}^{R} 2F_w(r,p,\cthet) r\dif{r}
\end{equation}
where 
\begin{equation}
a_i' = a_i \deriv{\mu}{\cthet} = a_i \frac{kp}{2(k-p)\epsmax},
\end{equation}
and the domain of integration has the following limits 
\begin{equation}
\pmin = k/2 \leq p_1 \leq p_2 \leq \pmax 
\end{equation}
with 
\begin{equation}
\pmax =  \frac{\epsmax k}{m} \frac{1 + \sqrt{1 - \frac{m^2}{\epsmax k} - \frac{\cthet m^2}{2 \epsmax^2} } }{2\epsmax/m + \cthet k/m},
\end{equation}
obtained by inverting eq. \eqref{eqmumin}. 
The limits for $\cthet$ are given by 
\begin{equation}
0 \leq \cthet \leq \cthet_{\max}
\end{equation}
with 
\begin{equation}
\label{eq:cmax}
\cthet_{\max} = 2 \left(\frac{\epsmax}{k} - \frac{m^2}{ k^2}\right),
\end{equation}
for which $\pmax = k/2 + \bigcirc(1)$.

In virtue of 37, $F_w$ now depends explicitly on $p$, hence the dependence in \eqref{eq:dwdc}. 

\newcommand{\cthetp}{{c_\theta'}}
For the lower-energy lepton, we need first to establish the kinematic relation between its outgoing angle defined by $\cos\theta' = 1 - \cthetp$ and the higher-energy lepton variables. All primed quantities refer to the lower-energy lepton. Taking the strong-photon direction along the $z$ axis we have the relation
\begin{equation}
1-\cthetp = \frac{p_z'}{p'}.
\end{equation}
Using the conservation of momentum \eqref{eqConstraintsLm}a) one can express $\cthetp$ to leading order
\newcommand{\pt}{\tilde{p}}
\newcommand{\mt}{\tilde{m}}
\begin{equation}
\label{eq:1mcp}
1 -\cthetp = \frac{1 - \frac{\pt}{1-\pt}\frac{\mt^2}{\pt^2} - \pt }{ \left(\left(1 - \pt - \frac{\mt^2}{2\pt}\right)^2 - \mt^2\right)^{1/2}} +\bigcirc(1)
\end{equation} 
where every quantities tilded quantity is expressed in unit of $k$, $\tilde{a} =a/k$.
Since there is no dependence on $\cthet$ one can directly deduce that 
\begin{equation}
\deriv{W_{\vect{k}}}{\cthetp} =\deriv{W_{\vect{k}}}{p} \abs{\deriv{p}{\cthetp} },
\end{equation}
which can be expressed using
\begin{equation}
\deriv{\cthetp}{p} = \frac{2\mt^2\left(\mt^4 + 2 (1-\pt)^2\pt(3\pt-1) + \mt^2 (1 - \pt(\pt(5-2\pt) + 2)\right)}{k(1-\pt)^2\left(\mt^4-4\mt^2\pt + 4(1-\pt)^2\pt^2\right)^{3/2}}
\end{equation}
after numerical inversion of \eqref{eq:1mcp}.
One finds that $\cthetp$ is a monotonously increasing function of $p$ and that 
\begin{eqnarray}
\label{eq:cpmin}
\cthetp(k/2) & = & 4\frac{m^2}{k^2}, \\
\cthetp(\pmax) & = & \frac{\epsmax^2}{m^2}\left(1 + \sqrt{1 - \frac{m^2}{k\epsmax}}\right).
\end{eqnarray}

\subsection{With conic boundary conditions}

\newcommand{\epsb}{\bar{\epsilon}}
\newcommand{\cx}{C_{\xi}}
\newcommand{\rb}{\bar{r}}
We consider the case where the soft photon distribution is defined everywhere between two cones of axis $\vect{k}$ and of half-apertures $0 \leq \xi_1 < \xi_2 \leq \pi$, and the distribution $F_w$ \eqref{eq:Fwdef} is given as a function of the coordinates $(\epsb=\epsilon/\epsmax, \cx=\cos\xi)$, where we deduce from \eqref{eqepsilon} and \eqref{eq:cosxi}
\begin{eqnarray}
\epsb & = & \frac{\rb^2}{4\mu} + \mu,\\
\cx & = & 1 - \frac{2\mu}{\epsb}  ,
\end{eqnarray}
where $\rb=r/\epsmax$.

We are now looking for the appropriate boundary conditions to apply to integral \eqref{eqWapproxsplitreduced}. Using the fact that
\begin{equation}
\tan\left(\frac{\pi}{2} - \xi \right) = \frac{z(r^2,\mu)}{r}
\end{equation}
where $z(r^2,\mu)$ is defined in eq. \eqref{eq:zdefintext}, one finds the new boundaries in $r$ by inverting this relation. The resulting $r$ boundaries are given by
\begin{equation}
r_{\xi_{1,2}}  = 2\epsmax \mu \left(\frac{1}{\tan\xi_{1,2}} + \frac{1}{\sin\xi_{1,2}}\right),
\end{equation}
where one checks that $r_{\xi_{2}} < r_{\xi_{1}} $. We need the intersection $[r_{\xi_{2}}, r_{\xi_{1}}]\cap[0,R_{\max}]$ which implies solving for $ R_{\max}(\mu_{\xi_{1,2}}) = r_{\xi_{1,2}}$, which gives us 
\begin{equation}
\mu_{\xi_{1,2}} =  \frac{1}{2} \frac{\sin^2\xi_{1,2}}{1 + \cos\xi_{1,2}}
\end{equation}
where one checks that $ \mu_{\xi_{2}} > \mu_{\xi_{1}}$. We can rewrite the energy spectrum \eqref{eqWfoutgoingleptonenergy} as

\newcommand{\muxia}{\mu_{\xi_1}}
\newcommand{\muxib}{\mu_{\xi_2}}
\newcommand{\pxia}{p_{\xi_1}}
\newcommand{\pxib}{p_{\xi_2}}
\newcommand{\rxia}{r_{\xi_1}}
\newcommand{\rxib}{r_{\xi_2}}

\begin{eqnarray}
\label{eq:dwdpconique}
& &\deriv{W_{\vectk} }{p}\left(\max\left(p, k-p \right)\right) = \\ 
&& \deuxpi cr_e^2\left( \int_{\mumin}^{\min\left(\max\left(\mumin, \muxia\right),1\right)} \dif{\mu} \int_{\max\left(0, \rxib\right)}^{\rxia}\dif{r} \: + \right. \nonumber\\
&& \left.  \int_{\min\left(\max\left(\mumin, \muxia\right),1\right)}^{\min\left(\muxib,1\right)} \dif{\mu} \int_{\max\left(0, \rxib\right)}^{R}\dif{r} \right) \sum_{i=1}^{4}  \frac{a_i}{\mu^i}  2  F_w(r, \mu)r. \nonumber
\end{eqnarray}

Concerning the angular spectrum, nothing more needs to be done for lower-energy leptons, and for higher-energy leptons we proceed similarly as for the energy spectrum above. Starting from  \eqref{eq:dwdc} one needs to replace $\mu$ by its expression as a function of $\cthet$ and $p$ in $R$. This allows us to define the $p$ analogs of $\mu_{\xi_{1,2}} $ by
\begin{equation}
p_{\xi_{1,2}} = k\frac{1 + \rb_{\xi_{1,2}}^2\frac{\epsmax}{\cthet k} + \sqrt{1-\rb_{\xi_{1,2}}^2}}{2 + \cthet\frac{k}{\epsmax} + \rb_{\xi_{1,2}}^2\frac{\epsmax}{\cthet k}},
\end{equation}
where one shows that $p_{\xi_{1}} <p_{\xi_{2}} $. The spectrum for higher-energy leptons is then obtained from \eqref{eq:dwdc}
\begin{eqnarray}
\label{eq:dwdcconique}
& &\deriv{W_{\vectk} }{c} = \\ 
&& c\deuxpi\left( \int_{\pmin}^{\min\left(\max\left(\pmin, \pxia\right),\pmax\right)} \dif{p} \int_{\max\left(0, \rxib\right)}^{\rxia}\dif{r} \: + \right. \nonumber\\
&& \left.  \int_{\min\left(\max\left(\pmin, \pxia\right),\pmax\right)}^{\min\left(\pxib,\pmax\right)} \dif{p} \int_{\max\left(0, \rxib\right)}^{R}\dif{r} \right) \sum_{i=1}^{4}  \frac{a_i'}{\mu^i}  2  F_w(r, \mu)r. \nonumber
\end{eqnarray}

The weak-photon distribution \eqref{eq:Fwdef} is defined by
\begin{equation} 
\label{eq_distribution_photons_faibles}
F_w(\epsb, \cx) = \sum_{n,m} F_w^{(n,m)} \epsb^n\cx^m.
\end{equation}

Integrations over $r$ in \eqref{eq:dwdpconique} and \eqref{eq:dwdcconique} yield expressions of the type 
\begin{equation}
\begin{array}{l}
\sum_{i=1}^{4}\frac{a_i}{\mu^i}\int_{r_1}^{r_2} 2F_w(r,\mu) r\dif{r} = \\
2\epsmax^2\sum_{i=}^{4} a_i\sum_{n,m}\sum_{l=0}^{m}\kparmin{m}{l}(-2)^{l} \mu^{l-i + 1} \left[\rb\epsb(\rb,\mu)^{n-l}\right]_{r_1}^{r_2}
\end{array}
\end{equation}
where
\begin{equation}
\label{eq:primitive_r_conique}
\left[\rb\epsb\left(\rb,\mu\right)^{n-l}\right]_{r_1}^{r_2} = \left\{ \begin{array}{ll}
2\mu\frac{\epsb\left(\rb_2,\mu\right)^{n-l+1} - \epsb\left(\rb_1,\mu\right)^{n-l+1}}{n-l+1} & \mathrm{if}\: n-l \neq -1\\
2\mu\log\left( \frac{\epsb\left(\rb_2,\mu\right)}{\epsb\left(\rb_1,\mu\right)}\right) & \mathrm{if}\:n-l=1
\end{array}\right. .
\end{equation}

To obtain the final spectrum \eqref{eq:dwdpconique} (resp. \eqref{eq:dwdcconique}), integration over $\mu$ (resp. over $p$) is possible analytically : the first line of \eqref{eq:primitive_r_conique} is a rational fraction that can be integrated throught partial fraction decomposition and the second line yields expressions of the type $\int x^k \log(\mathrm{polynomial}(x)) \dif{x}$ (where $k$ is integral)  which values are given in most relevant textbooks such as \citep{gradshteyn_table_2000}. However, the resulting expressions may be lengthy and a direct numerical integration might sometimes be more efficient.

\section{Applications}
\subsection{Isotropic black-body background distribution \label{sec:isotropic}}


\newcommand{\bb}{\mathrm{bb}}

\begin{figure*}
	\begin{center}
		\includegraphics[width = 0.8\textwidth]{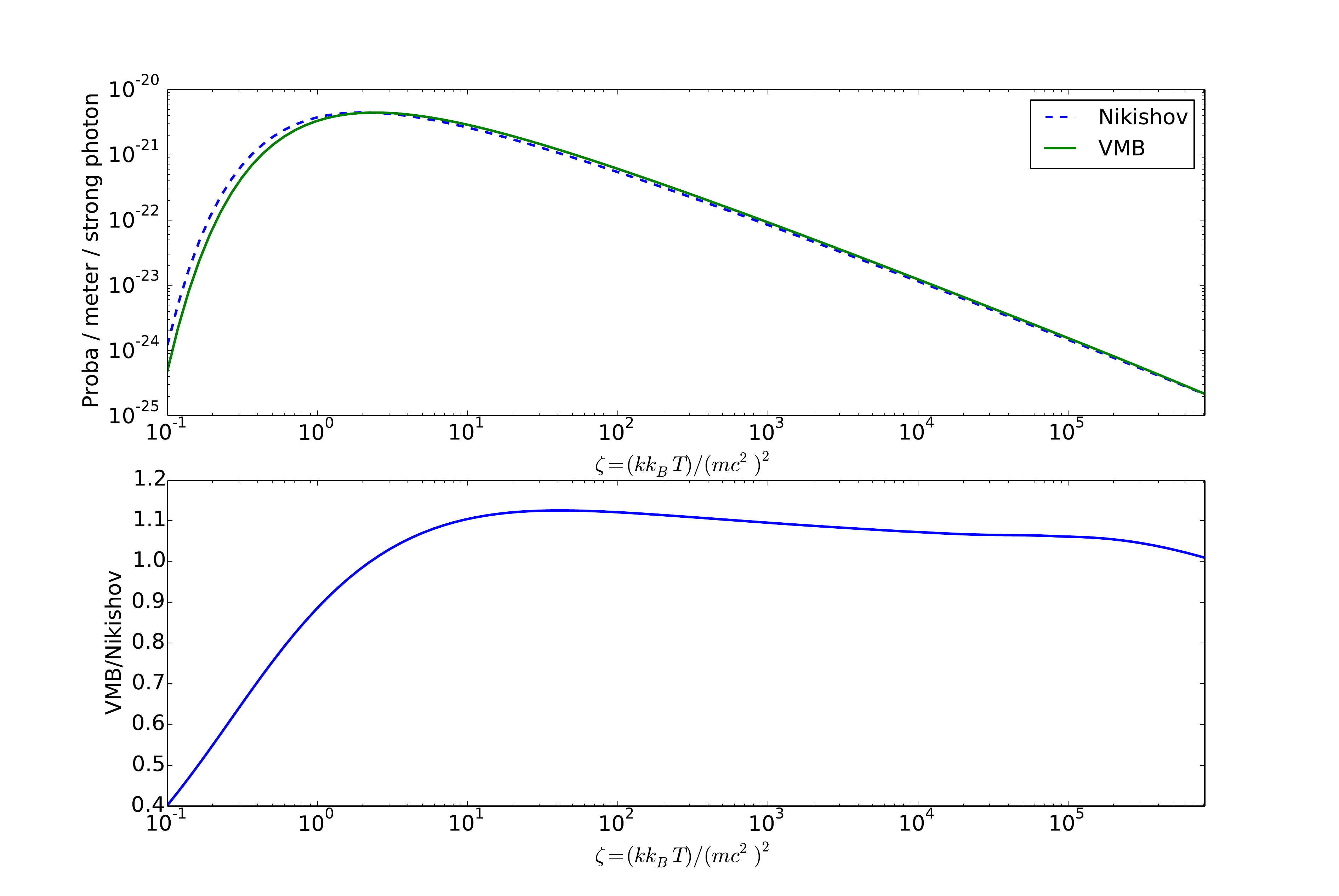}
	 \end{center}
	 \caption{\label{figcmb} Comparison of absorption of high-energy photons on a black-body background with Nikishov's formula (dashed line) and with our's (VMB, plain line). The scaling \eqref{eqscalefactor} is that of a black-body at $T_\bb = 2.7$ K (see formula \eqref{eqscalefactor}) to give an estimate of the effect of the cosmic-microwave background. In this case the energy of the strong photons ranges between $k \sim 100$ TeV and $k \sim 10^8$ TeV.  The bottom panel shows the ratio between the two theories . The ratio of probabilities averaged over $k$ is about $1.07$. The peak of our curve occurs around $2.6 m^2/T$ while Nikishov's is around $1.9 m^2/T$. The ratio between the two curves at the position of our peak is approximately $1.01$. }
%
%

\end{figure*}

\begin{figure*}
	\begin{center}
		\includegraphics[width = 0.8\textwidth]{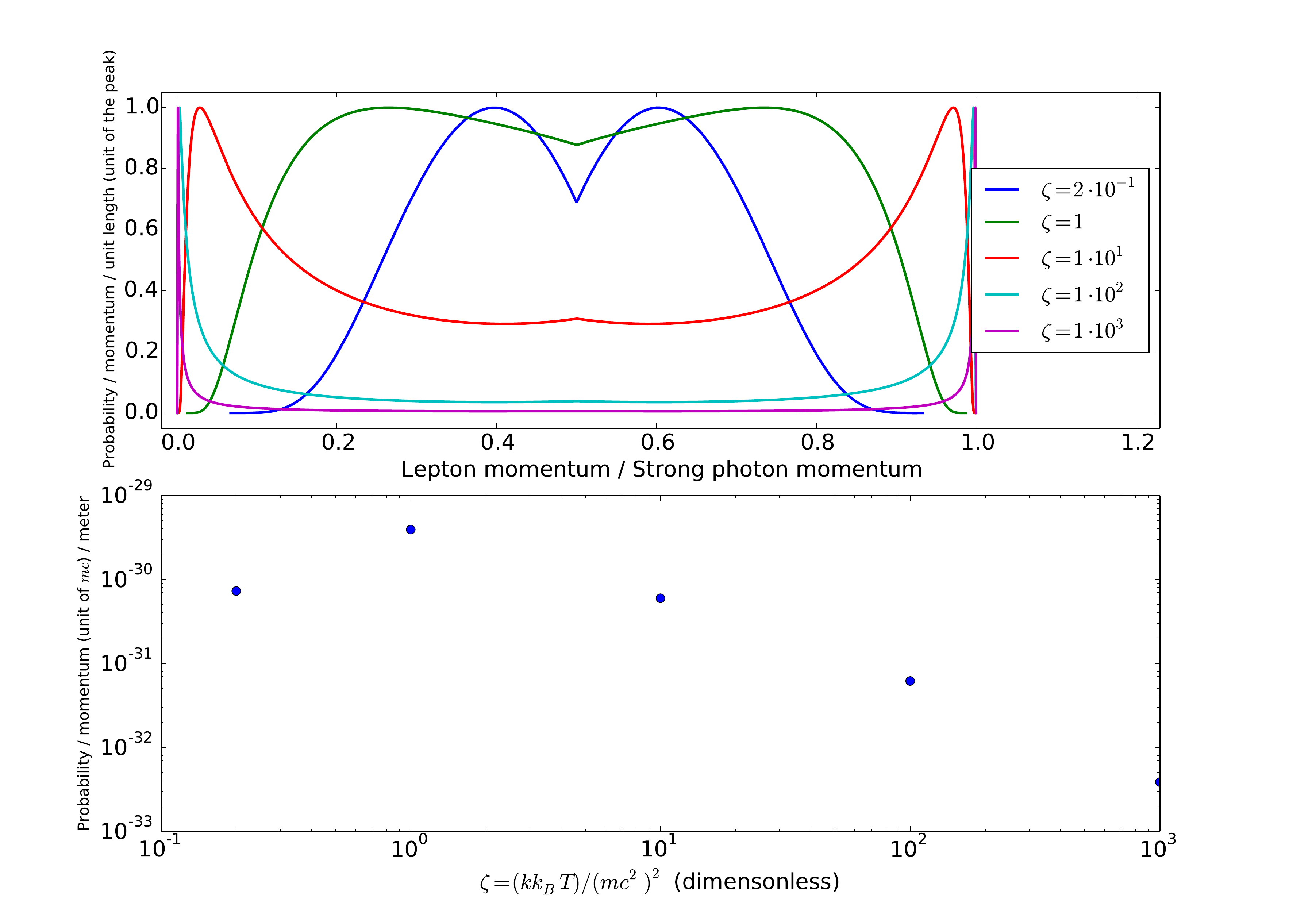}
	 \end{center}
	 \caption{ \label{figspectrumsbb} Spectra of outgoing leptons (electron or positron) for different strong-photon momenta $k$ on a black-body background at temperature $T_\text{bb}$ (top panel). $m$ is the mass of the electron, the speed of light and the Boltzmann constant are taken to be unity. The amplitudes are normalized to the amplitude of the peaks of each spectrum, and these amplitudes are reported on the lower panel. As in figure \ref{figcmb} these amplitudes are normalized to correspond to the cosmic-microwave background. The most intense peaks arise around a momentum $k$ such that its reaction with a background photon at $T_\text{bb}$ is at threshold, i.e. $k T_\text{bb} \sim m^2$. The more above threshold, the more separated, narrow and low the peaks are. The separation of the peaks can be understood as a mere  relativistic-frame effect, by analogy with a two-photon collision.}
\end{figure*}

Here we propose to check our approximation eq. \eqref{eqWapproxsplitreduced} against the exact isotropic case described in \cite{nikishov_1962,agaronyan_photoproduction_1983, boettcher_pair_1997}. We assume a high-energy photon hitting on a thermal soft photon background given by
\begin{equation}
\label{eqbbdistrib}
f_\bb(\epsilon) = \frac{2}{(\hbar c \deuxpi)^3}\frac{1}{e^{ \epsilon / k_B T} - 1}
\end{equation}
where $T$ is the temperature of the body and $k_B$ the Boltzmann constant. 
We choose a cutoff $\epsmax = 20 T$ (see Eq. \ref{mainapprox}) such that the neglected part of the black-body spectrum \eqref{eqbbdistrib} represents less than $\sim e^{-20}\sim 10^{-9}$ the total amount of background photons. We perform a Chebyshev interpolation (see e.g. \cite{grandclement_spectral_2009}) of $\epsilon^2 f_\bb(\epsilon)$ on the 25 first Chebyshev polynomials achieving a relative accuracy better than one thousandth everywhere and better than $10^{-6}$ for $0 \leq \epsilon/T \leq 10$, energies between which most of the photons are. This then allows us to derive the coefficients of the Laurent serie describing $f_\bb$ with poles of order one and two. Then, we produce the spectra of figure \ref{figspectrumsbb} and \ref{figcmb}.

On the top panel of figure \ref{figcmb} we plot the total probability of absorbing a strong-photon of energy $k$ as a function of 
\begin{equation}
\zeta = \frac{k k_b T}{(mc^2)^2}.
\end{equation} 
This parametrization by $\zeta$ makes the temperature dependency simple 
\begin{equation}
\label{eqscalefactor}
W_{\vect{k} } \propto \left(\frac{k_B T}{mc^2}\right)^3.
\end{equation} 
Here we choose to take $T = 2.7$K which allows to reproduce the result of  \cite{gould_opacity_1966} (dashed line) concerning absorption on the cosmic microwave background. 
The lower panel of figure \ref{figcmb} shows the ratio between our formula and the exact formula of \cite{nikishov_1962}. It shows that our result is fifty percent off at  $\zeta < 1$ and asymptotically tends to the correct value for large $\zeta$, the difference between the two curves is $\sim 10\%$ around the maximum of the curve located at $\zeta\sim2$. On average on the range plotted on fig.\ref{figcmb}, our formula overestimates the reaction rate by $7\%$.

A toy model can help us understand the shape of this curve. The peak of a black-body spectrum is roughly at $\epsilon_\mathrm{bb} \simeq 5 k_b T$. The cross-section peaks when the center-of-mass energy is $1.4(2m)$, so if one approximates the black-body spectrum to its peak one gets
\begin{equation}
\label{eq:seuiltoymodel}
\epsilon_\mathrm{bb} k(1-\cos\xi) \simeq 3.9 m^2. 
\end{equation}
For an isotropic distribution of soft photons, collisions take place at every angle $\xi\in[0,\pi]$. Taking the intermediate value $\xi = \pi/2$ one obtains from \eqref{eq:seuiltoymodel} an estimate of $\zeta$ for the peak of reactions
\begin{equation}
\zeta = \frac{k k_b T}{m^2} \simeq 0.8
\end{equation}
which is the right order of magnitude. One could argue that at such energies reactions would occur more face-on, meaning $\xi < \pi/2$ which is consistent with the higher peak position found on figure \ref{figcmb}. We now proceed to the computation of pair energy spectra.

Figure \ref{figspectrumsbb} shows the pair-creation spectra for different values of $\zeta$. Those spectra are directly computed using equation \eqref{eqWapproxsplitreduced} and  expressed as a function of $p/k$ which allows the same scaling law as in equation \eqref{eqscalefactor}, with $p$ the momentum of one of the created leptons, and normalize each spectrum to unity such that the obtained spectral shape are universal i.e. do not depend on the temperature of the black-body or on the absolute value of $k$, but only on $\zeta$. 
The shape and evolution of the spectra with the strong-photon energy is consistent with \cite{agaronyan_photoproduction_1983}. In this paper, the authors consider spectra resulting from the reaction of two isotropic monoenergetic photon distributions with energies $\epsilon$ and $k$ that are symmetrical with respect to $(k+\epsilon)/2$. Here, every spectrum is symmetrical with respect to $p/k = 0.5$ as result of neglecting $\bigcirc(\epsilon/k)$ terms.  Besides the shape of these spectra is very reminiscing of  pair-creation in the photon-plus-magnetic-field process that is well-known in the field of neutron-star magnetospheres \citep{daugherty_pair_1983}. The analogy is not fortuitous since the latter process can in principle be seen as the interaction of a strong photon with an assembly of magnetic-field photons. We see on figure \ref{figspectrumsbb} that each spectrum is made of two peaks that move apart and become narrower and weaker as the reaction occurs farther above threshold. Notice that the narrowing is relative to the momentum span and not absolute. 

The separation of the peaks at higher energies results from the fact that the cross-section favors alignment of ingoing and outgoing particles in the center-of-mass frame if the energy is much larger than the threshold energy. It follows that a Lorentz boost to the observer's frame along this axis results in a low-energy and a high-energy particle.  The intensity of the peaks of course depends on the background distribution, but also on the cross-section which decays as $\log(\tau)/\tau$ (see section \ref{sec_two_photon_general}). The latter dependency explains the above-threshold decrease of the peak intensity and the former explains the below-threshold decrease, as shown on the lower panel of figure \ref{figspectrumsbb}. 
One notices that spectra are not smooth in their center, which is naturally explained by our approximations that ensure continuity at the center but not continuity of derivatives.


\subsection{Above a hot neutron star\label{sec:pulsar}}

\begin{figure}
	\begin{center}
		\includegraphics[width = 0.4\textwidth]{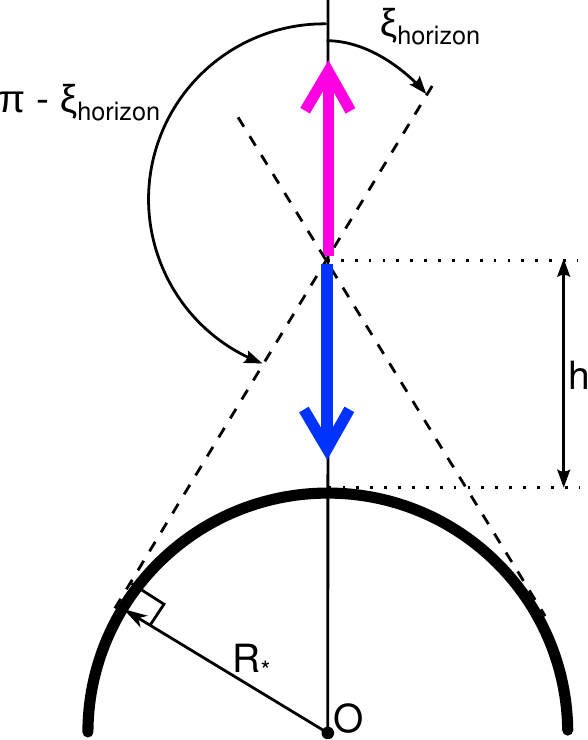}
	 \end{center}
	 \caption{\label{fig:neutronstar} A neutron star of center O and radius $R_*$ above which a up photon and a down photon are represented by radial arrows of opposite directions. Both photons are represented at an height $h$ above the surface of the star. From this height, they can interact with soft photons coming from the surface of the star within a cone of aperture $\xi_\mathrm{horizon}$, eq. \eqref{eq:horizon}, represented by dashed lines.  The incidence angle between the strong photon and soft photons therefore lies between $0$ and $\xi_\mathrm{horizon}$ for the up photon (purple upward arrow), and between $ \pi -  \xi_\mathrm{horizon}$ and $\pi$ for the down photon (blue downward arrow). }
\end{figure}

In this section, we consider a homogeneously hot neutron star  at temperature $T$ and two kinds of photons : the down photons and the up photons. Down photons are going radially toward the center of the star while up photons are going in the opposite direction, away from the star. This configuration aims at approximating a pulsar magnetic pole. Indeed, in a pulsar magnetosphere high-energy photons are expected to be mostly created by curvature radiation of electrons and positrons flowing along magnetic field lines that can be considered radial at low altitudes above the poles. Note that we do not consider only a hot cap here but the full star, as means of geometrical simplification.

\begin{figure*}
	\begin{center}
		\includegraphics[width = 1\textwidth]{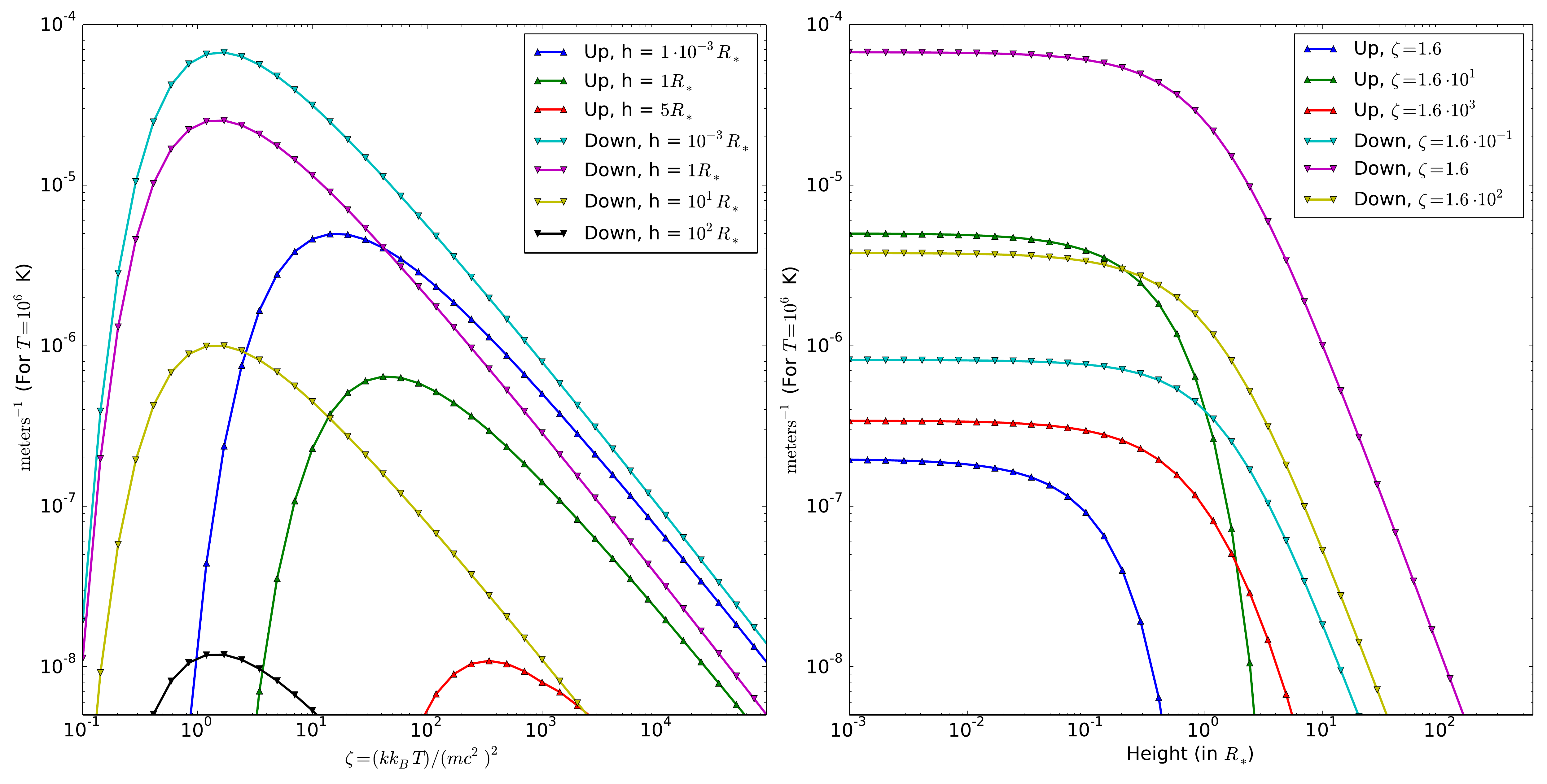}
	 \end{center}
	 \caption{\label{figpvsheight} Probability of reaction per meter of a strong photon of momentum $k$ as a function of $\zeta= (k k_B T)/mc^2)^2$ and height $h$ above a star of radius $R_*$. Up-triangle markers represent photons going radially up from the star. Down-triangle markers represent photons going down to the star. The probability scales like $T^3$, according to equation \eqref{eqscalefactor}, and is here represented using a fiducial $T= 10^6$K.
	 Left-hand-side panel : probability as a function of $\zeta$ at various heights.
	 Right-hand-side panel : probability as a function of height $h$ at various $\zeta$.  }

\end{figure*}

The case of pair production from photon-photon collisions in pulsar magnetospheres was studied by authors such as \cite{zhang_two-photon_1998, harding_regimes_2002}. In these papers, the authors generalize the formula of \cite{nikishov_1962} with a minimum energy threshold for the background distribution corrected by a factor $(1 - \cos\theta_c)^{-1}$ where $\theta_c$ is the maximum viewing angle on the hot polar cap of the star. In other words, they consider an isotropic black-body distribution where only photons within the viewing angle of the cap contribute, however with a threshold energy that corresponds to the largest incidence angle only since the threshold does not depend on the location of the emitter on the cap. Therefore, this approximation overestimates the threshold which generally translates in underestimating the reaction rate. This has little consequences when the viewing angle is wide, which is the case very close to the cap. However, one expects a faster decrease as one goes away from the cap and the factor $(1 - \cos\theta_c)^{-1}$ grows larger. As an example,  the authors of \cite{zhang_two-photon_1998} compute a maximum reaction probability of $5.7\dix{-5} \,  \mathrm{m}^{-1}$ at a viewing angle of $\SI{90}{\degree}$ when we get  $ 6.7\dix{-5} \SI{}{\meter^{-1}}$ (see peak of the down-photon $h=10^{-3}$ curve on figure \ref{figpvsheight} for an estimate), but they obtain only $ 6.3 \dix{-6} \,  \mathrm{m}^{-1}$ at $\SI{45}{\degree}$ when we still have a probability of $ 4.3  \dix{-5}  \, \mathrm{m}^{-1}$ (see their equation 9, for $T= 10^6$ Kelvins)). 

Besides, an interest of our formalism is that it can in principle deal with any  other orientation of the strong photon with respect to the star, and in particular the up photons.

In this configuration, the distribution of soft photons is still given by eq. \eqref{eqbbdistrib} except that it is now zero when the angle $\xi$ between the soft and the strong photon is beyond the horizon of the star as seen from the strong photon (see figure \ref{fig:neutronstar}). For a photon going upward, the horizon is defined by 
\begin{equation}
\label{eq:horizon}
\sin\xi < \frac{R_*}{R_* + h} = \sin\xi_{\mathrm{horizon}}
\end{equation}
where $R_*$ is the radius of the star (typically 10km) and $h$ is the height above its surface. Consequently, we use eqs. \eqref{eq:dwdpconique} and \eqref{eq:dwdcconique} with angles $ \xi_1 = 0, \xi_2 = \xi_{\mathrm{horizon}}$ for a up photon and  $ \xi_1 = \pi - \xi_{\mathrm{horizon}}, \xi_2 = \pi$ for a down photon. 
 
Figure \ref{figpvsheight} shows the probability of reaction per unit length (we will sometimes say "reaction rate") as a function of $\zeta$ at various heights $h$ above the cap (left panel), and as a function of $h$ at various $\zeta$ (right panel). As in the previous subsection, the temperature dependance is $ T^3$ for a given value of $\zeta$. All the figures in this section are made with a fiducial temperature of $10^6$K. With this value the conversion from $\zeta$ to $k$ is : $k \simeq 5.9\dix{3}\zeta mc^2$. At the lowest altitude we computed, $ h = 10^{-3}R_*$, the peak of the reaction rate is around $ \zeta = 1.6 $ for down photons and about an order of magnitude higher for up photons $\zeta \simeq 16$. This is a direct consequence of the threshold eq. \eqref{eq:seuiltoymodel} given the less favorable incidence angles of up photons. Another point is that the position of the maximum shifts to lower $\zeta$ as height increases for down photons, but to larger $\zeta$s for up photons. As can be seen on the right-hand-side panel, the reaction probability per unit length is fairly stable (within a factor of two) until $\sim 1R_*$, after which it decays very sharply. The decay is sharper as $\zeta$ increases for down photons and smoother for up photons, which explains the crossing between some curves on the right-hand-side panel. 

A qualitative reasoning explains these behaviors. For a low-energy down photon (i.e. $\zeta \lesssim 1$), most of the soft photons most likely to react are in a narrow cone almost face-on with the strong photon. The aperture of the cone defines the limit beyond which the reaction is below threshold. When the strong photon is higher, the almost face-on soft photons are the last to disappear because of the shrinking of the viewing angle. As energy rises, this cone becomes wider since soft photons provoking a near-threshold reaction are located at a wider angle according to formula \eqref{eq:seuiltoymodel}. Inside the first cone also appears a co-axial cone with a narrower aperture inside which photons are not contributing significantly anymore, since reactions are too far above threshold (and therefore the cross-section is too small) because of  small incidence angles. At large strong-photon energies ($\zeta \gg 1$), the soft photons close to the outer cone are the first to disappear when the viewing angle shrinks because of a larger height. This explains the faster decay of the reaction probability with $h$ for  larger $\zeta$s of down-photon curves on figure \ref{figpvsheight}.   The same kind of reasoning applies for up photons. Because soft photons are arriving "from behind", there is always an inner cone inside which the reactions are below threshold, and an outer cone limited by the angle beyond which the cross-section is too small if $\zeta \gg 1$ or the viewing angle if the strong-photon energy is small enough. The lower the energy of the strong photon the wider the outer cone and the most sensitive to viewing angle  the reaction rate is. That explains why, contrary to down photons, the reaction rate decays slower with altitude when $\zeta$ is larger on figure \ref{figpvsheight}. 
With this reasoning, one also understands why the energy of the reaction-rate peak (left panel) is quite stable at low altitudes and becomes smaller for down photons at high altitudes ($\gtrsim 1 R_*$) or larger for up photons.

\begin{figure*}
	\begin{center}
		\includegraphics[width = 0.8\textwidth]{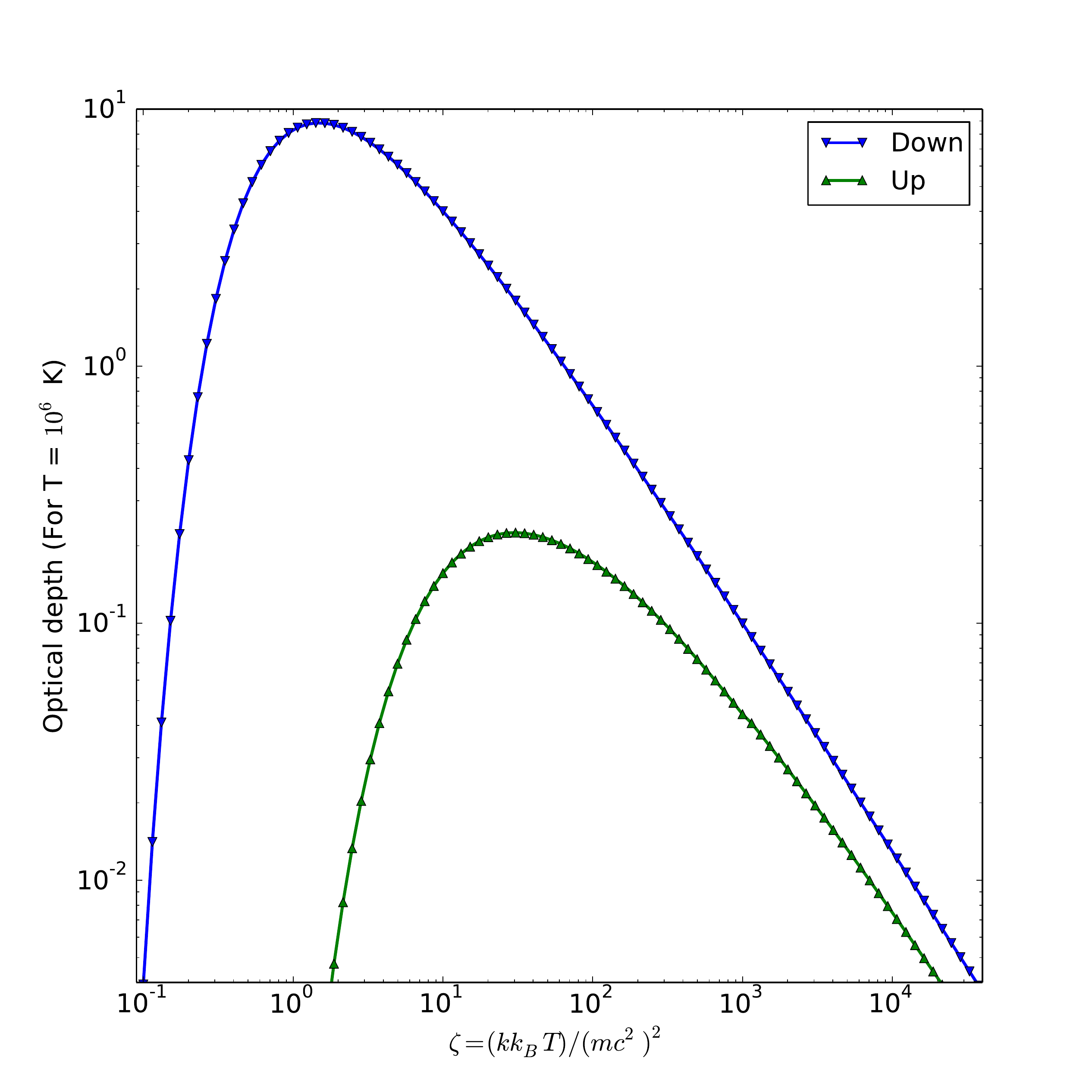}
	 \end{center}
	 \caption{\label{figoptdepth} Reaction rate in equation \eqref{eqWapprox1} integrated from $0$ to $10R_*$ for up and down photons as a function $\zeta$. The amplitudes are valid for a homogeneously hot star of temperature $T=10^6$K, and can be converted to other temperatures using the $T^3$ scaling law \eqref{eqscalefactor}. For photons going down toward the star, the peak is at $\zeta \simeq 1.4$ with an amplitude of $\simeq 8.8$. For photons going up away from the star, the peak is at $\zeta \simeq 30$ with an amplitude of $\simeq0.23$. }

\end{figure*}

Figure \ref{figoptdepth} shows the optical depth of strong photons as a function of $\zeta$ through $10 R_*$ from the surface. The optical depth is defined by
\begin{equation}
\tau_\zeta(10 R_*) = \int_0^{10 R_*} W_\zeta(h) \dif{h}.
\end{equation}
Because of the effects mentioned above the peak for down photons is slightly shifted downward at $\zeta \simeq 1.4$ while upward for up photons at $\zeta \simeq 30$. The corresponding typical Lorentz factors of the created particles are $8\dix{3}$ and $2\dix{5}$ respectively. The peak optical depths are respectively $\tau \simeq 8.8$ and $\tau \simeq 0.23$ at a temperature of $10^6$ Kelvins. One concludes that at this temperature more than three out of four up photons at the peak energy escape the magnetosphere if no other reaction or source of soft photons opacifies it. The magnetosphere may become opaque if the star is hotter than $\sim 1.6\dix{6}$K, temperature for which the maximum optical depth reaches $1$ owing to the $T^3$ dependence of the reaction rate.  Down photons with $\zeta$ between $\sim 0.25$ and $\sim 64$ have optical depths larger than unity and therefore are absorbed before they hit the star except if they are emitted at very low altitudes $h \ll R_*$.  The maximum optical depth of down photons is below one, namely the magnetosphere is transparent, for a temperature below $0.5\dix{6}$K.  Let's notice that our approximation of a uniformly hot star obviously leads to overestimating the optical depth on distances larger than the size of an actual hotspot.

\begin{figure*}
	\begin{center}
		\includegraphics[width = 0.8\textwidth]{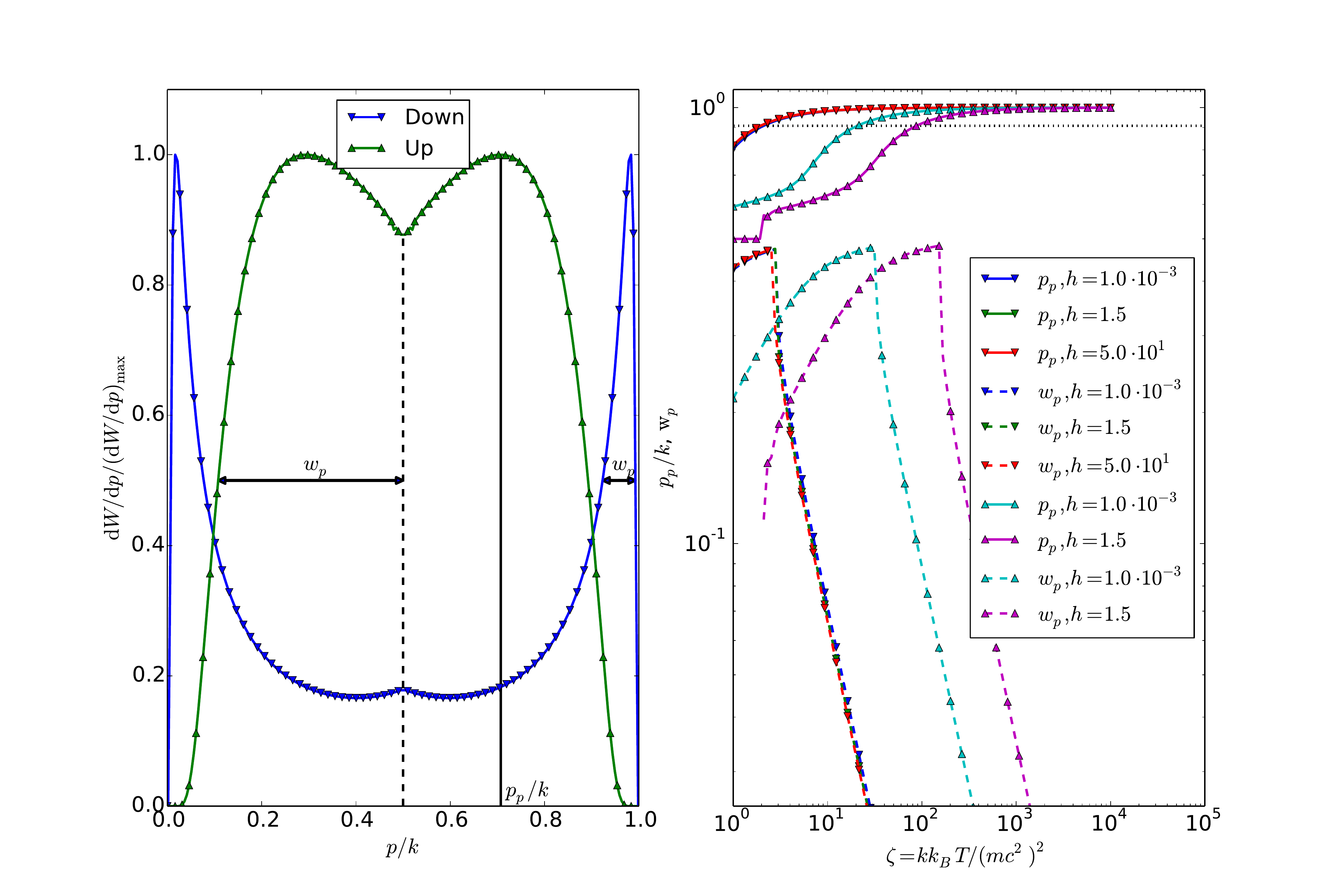}
	 \end{center}
	 \caption{\label{figdwdp} Left-hand-side panel : example of two normalized spectra of energy of created particles. These spectra are normalized to the amplitude of the largest peak, and the energy in abscissa is normalized to the energy of the incident strong photon  $k$. In both cases, $k$ corresponds to $\zeta = 10$ at an altitude $h = 0.5 R_*$, and the only difference resides in the up or down orientation of the strong photon. As in the isotropic case \ref{figspectrumsbb}, spectra are generally made of two peaks more or less thin and separated. The width at half maximum of peaks $w_p$ is defined in the two possible cases : if one side of a peak never reaches its half before rising again to another peak in which case the width is taken to be half of the double-peak width, or if the peak is well defined on both sides in which case the definition is straightforward. The position of the most energetic peak $p_p/k$ is defined as well.
	 	 Right-hand-side panel : Evolution with $\zeta$ of positions $p_p/k$ of the higher-energy peak (curves on the higher part of the plot), and widths at half maximum $w_p$ (curves on the lower part of the plot) for up and down photons at various heights $h$ (in units of $R_*$). Positions are ranging from $0.5$ at low $\zeta$s which corresponds to a perfectly centered peak or to a null spectrum when a reaction is below threshold (lowest energies of up photons), to $\simeq 0.98$ at large $\zeta$s. The horizontal dotted line shows the positions at which the ratios between the two peaks is $10$. Widths at half maximum are rising to $\sim 0.45$ until the two peaks separate and drop sharply to $\simeq 0.029$.}

\end{figure*}

Figure \ref{figdwdp} shows the energy spectra of the created leptons (left panel) and the evolution of the position and widths of the peaks as a function of $\zeta$ at various heights (right panel). The spectra have the same double-peaked structure as in the isotropic case (figure \ref{figspectrumsbb} ) but evolve differently depending on the orientation of the strong photon. The general principle is the same : the more above threshold the more separated peaks, with the consequence that they narrow when they get close to the limits of $p/k \in [0,1]$. For down photons, the width of the peaks $w_p$ (in unit of $k$) has very little dependence on altitudes which is due to the fact that for the range of $\zeta \lesssim 20$ visible on this plot (right panel), the efficient soft photons are mostly face-on and suffer no effect of viewing angle. The same thing applies for the position of the most energetic peak $p_p$ (and the least energetic at $k-p_p$). Down-photon peaks are wide $w_p\sim 0.45$ for $ \zeta \lesssim 2$ and then sharply narrow while their position smoothly goes from $p_p/k \sim 0.8$ to $p_p/k \lesssim 1$  at large $\zeta$s. On the contrary, up photons are very sensitive to altitude, which is explained by the fact that the higher above the star, the narrower the viewing angle and therefore the incidence angle, and the more energetic up photons need to be for the reaction to be at or above threshold. As a consequence up-photons peaks are very centered at low values of $\zeta$s, with  $p_p/k \sim 0.6$, and are even more centered at higher altitudes.  With $\zeta$ rising, the energy distribution becomes increasingly asymmetric as $p_p/k \rightarrow 1^-$, although it takes a larger $\zeta$ at higher altitude. Similarly, peak widths are growing with $\zeta$ until a maximum $w_p\sim 0.45$ at a $\zeta$ all the more large that altitude is high, after which $w_p$ drops sharply. This sharp change of slope happens because the two peaks separate (see comment of figure \ref{figdwdp} ).

\begin{figure*}
	\begin{center}
		\includegraphics[width = 0.8\textwidth]{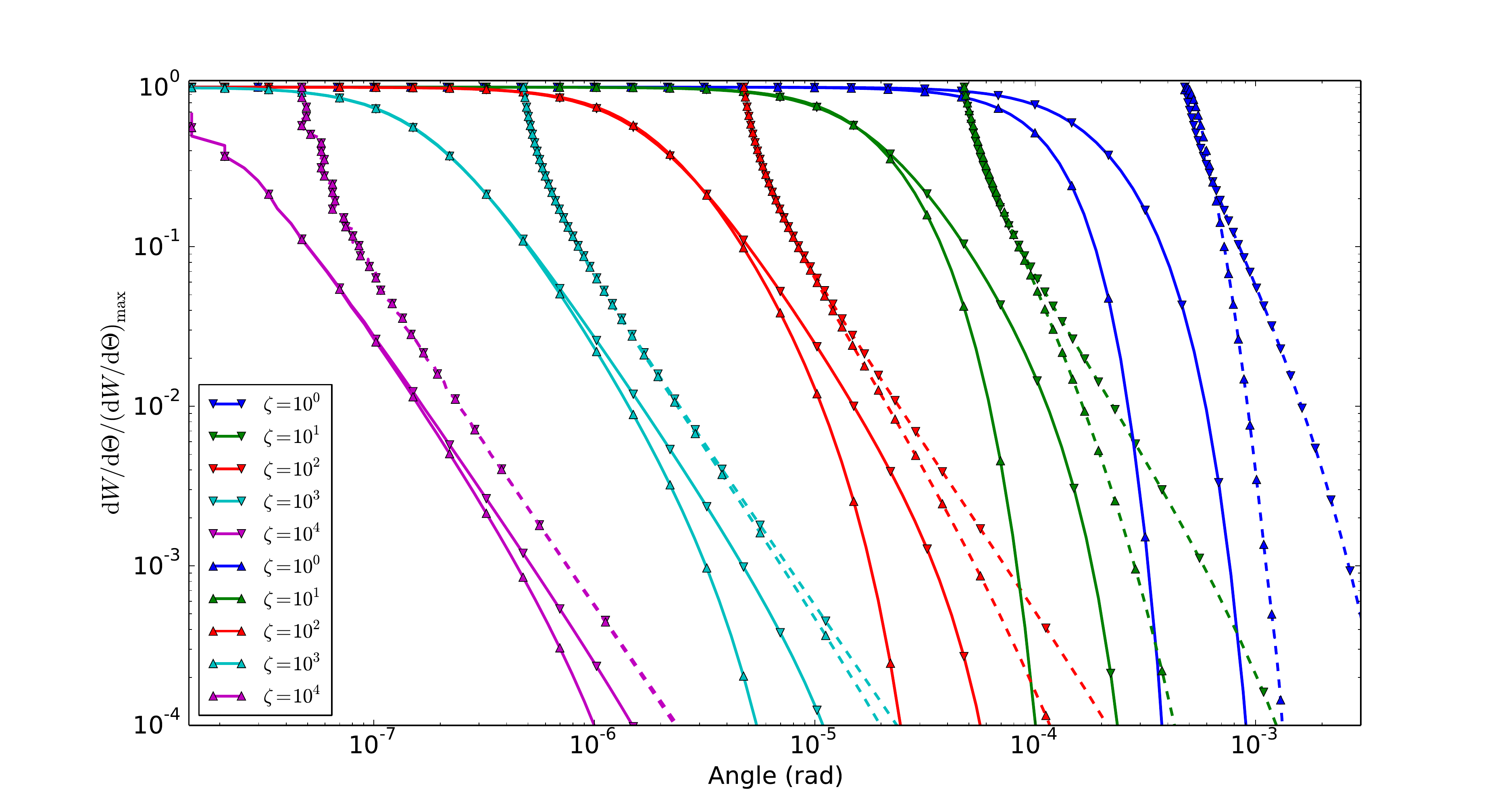}
	 \end{center}
	 \caption{\label{figdwdthet} Angular spectra for various values of $\zeta = \frac{k k_b T}{(mc^2)^2}$ and both up and down orientations (respectively up-triangle and down-triangle markers) normalized to their maximum values. Solid lines correspond to angular spectra of the higher energy leptons ($ p > k/2$) and dashed-line spectra to their lower-energy counterparts ($p < k/2$). All these spectra correspond to a height above the star $h= 0.001R_*$ which is representative for all other heights. Indeed their amplitudes significantly change with height but their shapes (and therefore the shown normalized spectra) barely change.   }
\end{figure*}

Figure \ref{figdwdthet} shows the normalized angular spectra for both up and down photons, and both higher-energy ($p > k/2$) and lower-energy (p < $k/2$) outgoing leptons at various values of $\zeta$. It is remarkable that apart from their amplitudes (not visible on this normalized plot), these spectra do not change much with height apart at large and very unlikely angles, and therefore we limit ourselves to only one height. These spectra are monotonously decreasing as the angle becomes larger, and the larger $\zeta$ the larger the outgoing angle. Lower-energy leptons have larger outgoing angles than their higher-energy counterpart and are not created below a minimum angle defined in equation \eqref{eq:cpmin}. For a given $\zeta$, leptons created from down photons are always going out at larger angles, and the difference is growing at larger angles of the spectrum. In a pulsar magnetosphere the outgoing may be important because the pairs will radiate more or less synchrotron radiation depending on their momentum perpendicular to the local magnetic field. We see here that the angles with respect to the progenitor strong photon are overall very small, which is expected from relativistic collimation. If one assumes that strong photons are produced though curvature radiation along the magnetic-field lines, then the angle distributions presented on figure \ref{figdwdthet} matter only if the mean free path is much shorter than the radius of curvature of the field line. This is not the case with the parameters presented in this section, and would probably require an extra source of photons. 

\section{Discussion}

Recent simulations of aligned millisecond-pulsar magnetospheres indicate that significant pair production may occur near the so-called separatrix gap and y point (see \cite{cerutti_electrodynamics_2016} and references therein) near the light cylinder. This implies that the source of pairs be photon-photon collisions. However, in the most detailed modeling of pair creation realized by \cite{Chen_Beloborodov_2014}, photon-photon pairs are created with a constant and uniform mean free path of $2R_*$. If one assumes that the source of soft photons is only provided by the star, this assumption seems reasonable close to the star, $h < 2R_*$, but greatly underestimated beyond owing to the exponential cutoff of the reaction rate with altitude (figure \ref{figpvsheight}). This issue can be overcome if another source of soft photons can be found, resulting for example from synchrotron radiation near the light cylinder. Moreover, in these simulations, the direction of strong photons relative to the soft-photon sources is not taken into account, which can have an effect of several orders of magnitude on reaction rates with a strong dependence on strong-photon energies (see figure \ref{figoptdepth}). The energy separation of the two outgoing leptons (figure \ref{figdwdp}) may also have an important impact on the subsequent synchrotron radiated by the pair. Indeed, the synchrotron peak frequency scales like $\gamma^2$, where $\gamma$ is the Lorentz factor of the particle around the magnetic field. Therefore, a typical situation in which the higher-energy lepton takes $10$ times more energy than the other (dotted line on figure \ref{figpvsheight}) results in two synchrotron peaks radiated two orders of magnitude apart. This situation is reached at values of $\zeta$  for which the optical depth on figure \ref{figoptdepth} is still high i.e. more than half the peak value. Notice that we implicitly assume here that both particles share the same angle with respect to the local magnetic field, which is justified by small outgoing angles shown on figure \ref{figdwdthet}. 

\begin{figure*}
\begin{center}
\includegraphics[width = \textwidth]{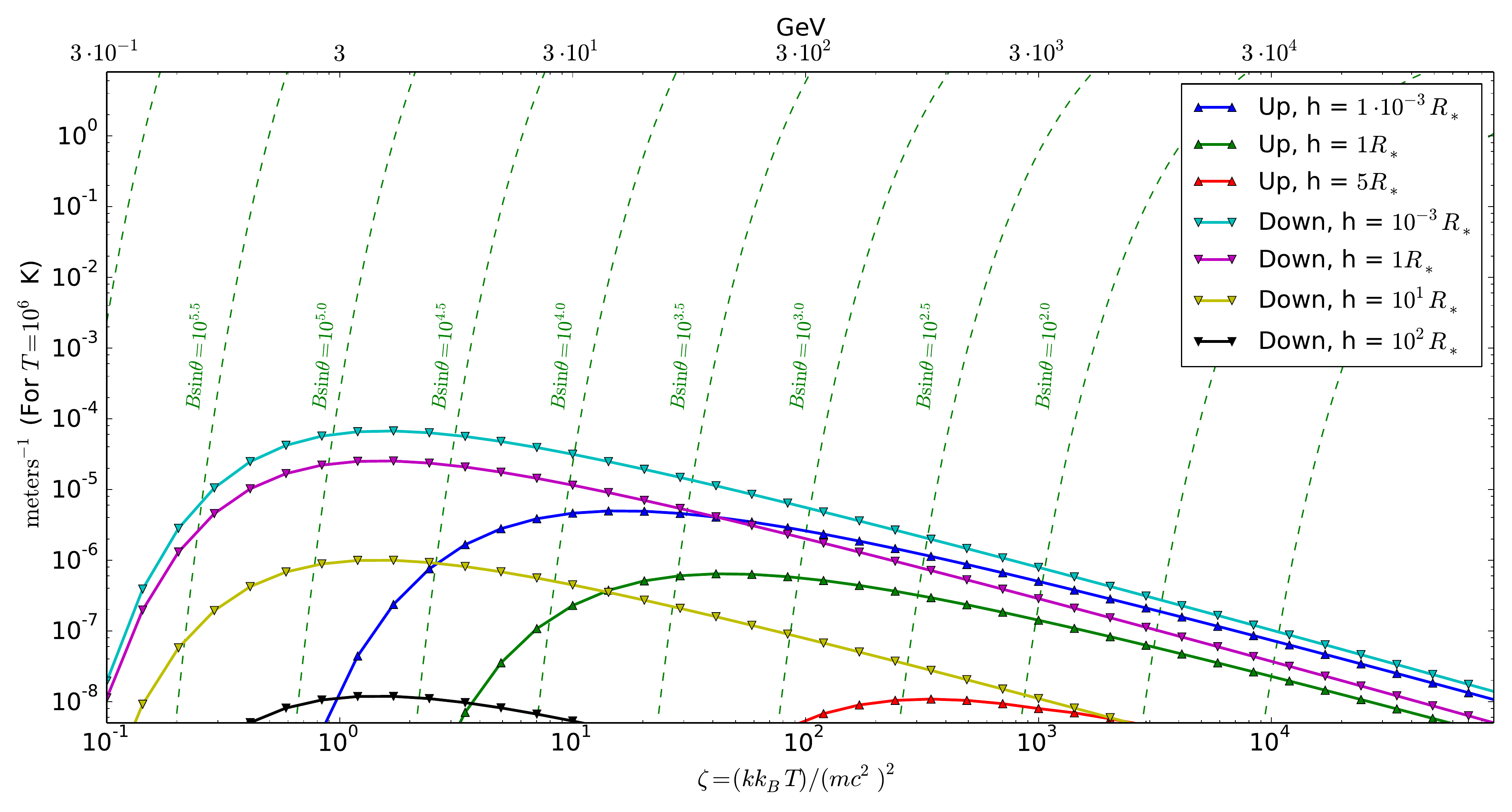}
\end{center}
\caption{\label{fig:comparisonggvsBg}Comparison of reaction rates of the $\gamma\gamma\rightarrow e_+ e_-$ process (solid lines) versus the $ \gamma + \vec{B} \rightarrow e_+e_-$ process (dashed lines). The photon-photon reaction rates are identical to those on the left-hand-side panel of figure \ref{figpvsheight}, but a temperature of $T=10^6K$ is taken giving the photon energies in GeV reported on the upper horizontal axis corresponding to the $\zeta$ values on the lower axis. The photon-magnetic-field reaction rates are given in the $\chi \ll 1$ approximation (see text) within which they depend only on the the photon energy and the magnetic intensity perpendicular to the photon direction $B\sin\theta$ (in Teslas), where $\theta$ is the angle between the local magnetic field and the photon direction. }
\end{figure*}

The photon-photon mechanism competes with the photon-magnetic-field mechanism $\gamma + \vec{B} \rightarrow e_+ e_-$ for the creation of pairs in pulsar magnetospheres. For our present discussion, we focus on magnetic fields smaller than the critical magnetic field $B_c = 4.4\dix{9}$ Teslas. Additionally the photon energies produced by curvature or synchrocurvature radiations in pulsar magnetospheres cannot exceed $\sim 100$ GeV owing to radiation reaction (see e.g. \citet{vigano_compact_2014}). With these two limits, the photon-magnetic-field reaction rate can be computed with the asymptotic expression \citep{tsai_propagation_1975, daugherty_pair_1983} 
\begin{equation}
\label{eq:gBrr}
W_{\gamma \vec{B}} \underset{\chi \ll 1}{\simeq} 4.3\dix{9} \frac{B\sin\theta}{B_c} \exp\left(-\frac{4}{3\chi}\right) \mathrm{m}^{-1} \:\: \mathrm{with} \:\: \chi = \frac{\hbar\omega}{2mc^2}\frac{B\sin\theta}{B_c},
\end{equation}
where $\hbar\omega$ is the energy of the gamma photon, $B$ the intensity of the local magnetic field, and $\theta$ the angle between the direction of the magnetic field and the direction of the photon. The photon-magnetic-field optical depth heavily depends on the magnetic-field geometry: typically, a photon in the pulsar magnetosphere is emitted parallel to the local magnetic field due to relativistic beaming, thus starting with a reaction rate $W_{\gamma \vec{B}} = 0$ which increases along the propagation as $\theta$ increases. The upper limit of the reaction rate of the photon can be estimated by considering $\sin\theta =1$ everywhere, although this is bound to largely overestimate the probability of creating a pair.   

Figure \ref{fig:comparisonggvsBg} shows a comparison between the reaction rate of equation \eqref{eq:gBrr} for a range of values of $B\sin\theta$ versus the photon-photon reaction rates computed at various altitudes of figure \ref{figpvsheight}. Considering that the range of probable photon energies lies below $100$ GeV one sees that the photon-photon mechanism can dominate near the surface of millisecond pulsars where $B\sin\theta \lesssim 10^5$ Teslas, and clearly dominates for down photons at altitudes $h\geq 10R_*$ where, assuming a dipolar magnetic field where $B\propto (R_* + h)^{-3}$, one has $B\sin\theta \lesssim 100$ Teslas for millisecond pulsars. For up photons, it is less clear which mechanism dominates without taking into account a particular magnetic geometry. It should also be noted that, if a comparable soft-photon density can be achieved in the outer magnetosphere as is necessary in some recent simulations (e.g. \citet{Chen_Beloborodov_2014}), the photon-photon mechanism would be largely dominating the photon-magnetic-field mechanism in this region of the magnetosphere.

We have neglected two effects possibly important in our application to a hot neutron star, section \ref{sec:pulsar}: general relativity and the effect of the strong magnetic field on the pair-production cross section. The former effect, general relativity, redshifts the spectrum of soft photons, and enlarges the visible horizon of the star. Indeed, light bending due to the gravitational field of the star curves the trajectories of soft photons in such a way that part of the surface beyond the geometrical horizon of the star becomes ``visible'' by the strong photon. Simple analytical formulas for the effective soft-photon distribution with general-relativistic effects have been given by \citet{beloborodov_gravitational_2002, turolla_pulse_2013}. However, these expression are valid for an observer (in our case the strong photon) located at infinity and are therefore inapplicable in the present case where we consider reactions between $10^{-3}R_*$ and $100R_*$. Instead, one would need to compute numerically the geodesics followed by the soft photons between the surface and the strong photon. 

The second neglected effect is the effect of a strong magnetic field on the cross section for photon-photon pair creation. It has been worked out by \citet{kozlenkov_two-photon_1986}. This cross section shows a sawtooth behavior at energies corresponding to the quantified Landau levels of the outgoing leptons. Unfortunately, this cross section is very unwieldy for practical calculations. However, its effect is low or moderate in magnetic fields much lower than the critical field $B_c$, which fortunately corresponds to the range of parameters where photon-photon pair creation can dominate over photon-magnetic-field pair production (see above and figure \ref{fig:comparisonggvsBg}), and therefore the range of interest of our formalism. As mentioned before, the other important domain of application of our formalism is the outer magnetosphere where the magnetic field is also much smaller than $B_c$, including in the case of young pulsars such as the Crab. 

\section{Conclusion}

We propose a formalism to analyze photon-photon pair creations with an arbitrarily anisotropic soft-photon background. This formalism  allows to calculate energy and angle spectra of outgoing pairs, as given by formulas \eqref{eq:dwdpconique} and \eqref{eq:dwdcconique} respectively.

 Calculations are carried using two approximations : the first being that the strong photon is much more energetic than the soft-photon cutoff  energy $\epsmax$, and the second that the outgoing higher-energy lepton of momentum $\vect{p}$ be very aligned with the progenitor strong photon of momentum $\vect{k}$ in the sense that $\left(\vect{k} - \vect{p}\right)_{\perp} / \left(\vect{k} - \vect{p}\right)_{\parallel} \ll 1$, \eqref{dominpara}, where perpendicular and parallel components are taken with respect to $\vect{k}$.  This latter approximation is the most stringent one. Indeed,  one can show that the inequality itself ($<1$) is always true within the frame of our first approximation, but its large validity $(\ll 1)$ comes if the reaction is far above threshold i.e. $k\epsmax/m^2 \gg 1$.
 
In section \ref{sec:isotropic}, we compare our formalism with the exact formula that can be found in the literature (\cite{nikishov_1962}, or \cite{agaronyan_photoproduction_1983} eq. 4 and 5 for a more detailed formulation), and show that our approximated formulation gives results accurate at $\sim 7\%$ on average, with $\sim 10\%$ near the peak and asymptotically tend to the exact value at high energies. However, the difference can be as large as $\sim 50\%$ at low energies. We show pair spectra that are consistent with those of \cite{agaronyan_photoproduction_1983} in the isotropic case. 

In section \ref{sec:pulsar} we show that the differences created by the strong anisotropy of radiation near a hot neutron star are much more important than a few percent, potentially reaching several orders of magnitude depending on energy, direction of the strong photon, and altitude above the star. We consider two directions for strong photons : radially toward the star (down photons) and away from the star (up photons). In both cases reaction rates are stable until $1R_*$, before undergoing an exponential cutoff. However, the peak of strong-photon absorption occurs at an energy $\sim 10$ times larger for up photons. Energy pair spectra show two peaks symmetric with respect to $k/2$, similarly to the isotropic case. These peaks separate as the energy of the reaction rises. We show that such a difference in energy between the two outgoing leptons can importantly affect the synchrotron emission of the pairs for a large range of strong-photon energy compared to a simple model in which both components of a pair take away the same energy. 

These findings are meant to contribute to a better modeling of pair creation from photon-photon collisions in pulsar magnetospheres. Recent millisecond-pulsar-magnetosphere simulations gave an important role to this pair-production mechanism \citep{cerutti_electrodynamics_2016}. However, the current state of modeling leaves an important uncertainty on the amount of soft photons needed to sustain such pair discharges. The results of this work provide means to estimate the mean free path on a soft-photon background resulting from a homogeneously hot neutron star. Moreover it provides formulas to obtain results with virtually any soft photon distribution, in particular resulting from secondary synchrotron close to the light cylinder. The possibility to generate energy spectra allows to differentiate  between the two components of a pair and therefore to differentiate their synchrotron emissions.




\bibliographystyle{mnras}
\bibliography{pairesggee} 



\appendix

\section{Derivation of the general result \label{sec_correct_rate}}
\label{sec_correct_rate}
We show the that domain of integration can be approximated by an hyperboloid of revolution. Then, we compute the integral $W_k$ over this surface assuming that the distribution function of weak photons is given by a polynomial (e.g. Taylor expansion). A variety of notations and relations is used, we summarize them in appendix \ref{apformulae}.

\subsection{Parametrization of $L_-$ by the three-momentum of the weak photons} 



The spectrum of pair creation is the density of probability of making a pair as a function of the energy of one of the particles. By definition, it is symmetric with respect to half of the total energy $k + \epsilon \simeq k$: if one of the particles has an energy $p$ then the other has $k-p$ as a result of energy conservation. Therefore we consider only the upper half of the spectrum, for $p > k/2$ and
\begin{equation}
\label{eqWsymmetry}
\deriv{W}{p}(p) = \deriv{W}{p}(k-p)
\end{equation}
Therefore, we are left with the very helpful ordering
\begin{equation}
\label{eqapproxp}
k \gtrsim p \gg  m, \epsmax,
\end{equation}
which allows to write :
\begin{equation}
	P^0 = p + \frac{m^2}{2p} + \bigcirc\left(\frac{m^2}{p^2}\right).
\end{equation}
Further, we learn from the angle-averaged cross-section \citep{berestetskii_quantum_1982} that when the reaction is way above threshold, one of the particles of the pair takes most of the energy while the other takes almost nothing (section \ref{sec_two_photon}), which reinforces our assumption. 
In the following calculation we note $\bigcirc\left(n\right)$ a development up to a bounded function of 
\begin{equation}
\label{eqbigon}
\left(\frac{m}{k}\right)^n \sim \left(\frac{m}{p}\right)^n \sim \left(\frac{\epsmax}{k}\right)^n \sim \left(\frac{\epsmax}{p}\right)^n.
\end{equation}
\begin{figure}
	\begin{center}
		\def\svgwidth{0.2\textwidth}
	    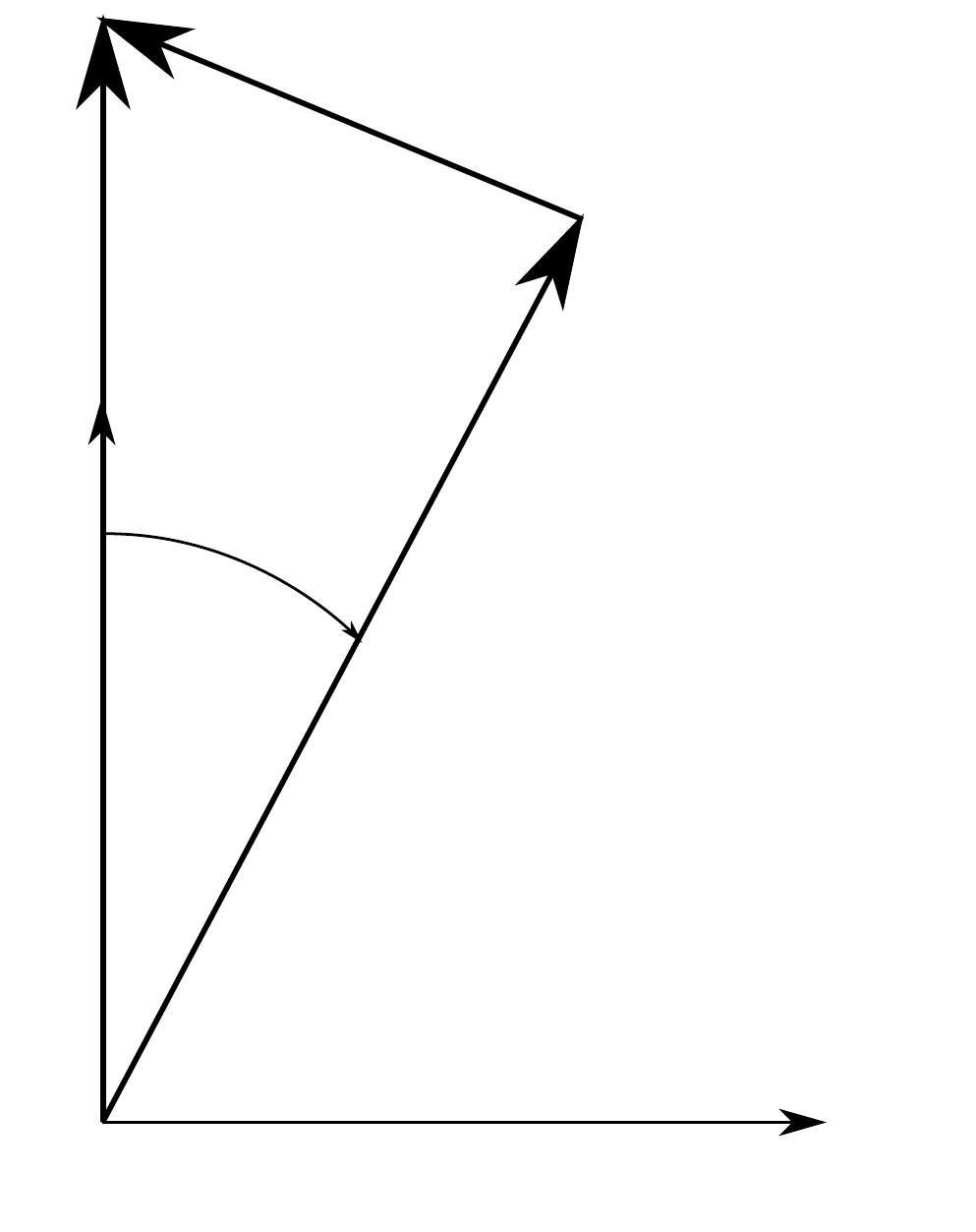
	 \end{center}
	 \caption{\label{approxcoord} Coordinate system. In our approximation, $\vect{k}$ is a quasi symmetry axis. }
\end{figure}
This leads to the conclusion that $\vect{p}$ is almost aligned with $\vect{k}$. Indeed, we can show that
$\epsilon_{min} < \epsilon_{max}$ implies that momentum can be conserved only if $\cos\theta > C_\text{min}$, where 
\begin{eqnarray}
C_\text{min} &=& \frac{k+\epsmax}{kp}(P^0 - \epsmax) \\ \nonumber 
&-& \frac{\epsmax}{kp}  \sqrt{k^2 + p^2 - 2kP^0 + 2k\epsmax - 2P^0 \epsmax + \epsmax^2}.
\end{eqnarray} 
This is approximated as
\begin{equation}
C_\text{min} =1 + \frac{m^2}{2p^2} - 2 \frac{\epsmax(k - p)}{kp} + \gdo{3}.
\end{equation} 
Therefore we set $\vect{k}$ as the main axis of our coordinate system (Figure \ref{approxcoord}), parallel to the unit vector $\vz$ of the direct triad $(\vx, \vy, \vz)$. 
For a weak photon of energy $\epsilon$ :
\begin{equation}
 1 - \cos\theta \leq  2 \frac{\epsilon(k - p)}{kp} - \frac{m^2}{2p^2} + \gdo{3}
\end{equation}
By squaring relevantly the mass-shell constrain \ref{eqConstraintsLm}b) one obtains the following quadratric constrain :
\begin{equation}
\label{eqsquareshell}
 \left(\epsilon \left(P^0 - k\right)\right)^2 = \left( \vect{x} \cdot \left(\vect{k} - \vect{p}\right)  + K\cdot P \right)^2
\end{equation}
which can be rewritten as 
\begin{equation}
\label{eqsquareshell1}
\vect{x}(\alpha^2 1 -  \overline{\beta}) \vect{x} - 2A\vect{\beta}\cdot\vect{x} = A^2,
\end{equation}
where
\begin{eqnarray}
A & = & K\cdot ,P \\
\vect{\beta} & = & \vect{k} - \vect{p}, \\
\alpha & = & k - P^0, 
\end{eqnarray}
and $\overline{\beta}$ is defined by
\begin{equation}
	\vect{x}\overline{\beta}\vect{x} = \left(\vect{\beta}\cdot\vect{x}\right)^2.
\end{equation}
We find 
\begin{eqnarray}
\overline{\beta} & = & \left( \begin{array}{ccc}
\beta_x^2 & 0 & 0 \\
2\beta_x\beta_y & \beta_y^2 & 0 \\
2\beta_x\beta_z & 2\beta_y\beta_z & \beta_z^2
\end{array}\right).
\end{eqnarray}
Let's rewrite Eq. (\ref{eqsquareshell1}) in a dimensionless form,
\begin{equation}
\vect{x}(\frac{\alpha^2}{A^2} 1 -  \frac{\overline{\beta}}{A^2}) \vect{x} - 2\frac{\vect{\beta}}{A}\cdot\vect{x} = 1.
\end{equation} 
The three proper values of this quadratic form are
\begin{equation}
\label{propervalues}
 (\frac{\alpha^2}{A^2} - \frac{\beta_x^2}{A^2}, \frac{\alpha^2}{A^2} - \frac{\beta_y^2}{A^2}, \frac{\alpha^2}{A^2} - \frac{\beta_z^2}{A^2} ).
\end{equation}
The geometrical type of this quadratic form is determined by the signs of its proper values. For this we express the different quantities using the approximation defined in Eq. (\ref{eqapproxp}),
\begin{eqnarray}
A & = & m^2\left(\frac{1}{2}\frac{k}{p} + \frac{kp}{m^2}(1 - \cos\theta) \right)+ \gdo{1}\\ \nonumber
  & = & 2m^2 \frac{k-p}{m}\frac{\epsmax}{m}\mu\\
\alpha & = &   m\left( \frac{k}{m}-\frac{p}{m} - \frac{m}{2p} \right) + \gdo{2}\\
\sqrt{\beta_x^2 + \beta_y^2} = \beta_\perp & = & p\sqrt{2(1 - \cos\theta)} + \gdo{3} \\
\beta_z = \beta_\|  & = & p\left(\frac{k}{p}  - \cos\theta \right)
\end{eqnarray}
%
It can be shown that, provided that $\epsmax < \frac{3}{8}m $ and for any relevant $\theta$ or $p$,
\begin{equation}
\label{domialpha}
\alpha > \beta_\perp
\end{equation}
Similarly,
provided that $\epsmax < \frac{1}{4}m $ (see Eq. \eqref{mainapprox}),
\begin{equation}
\label{dominpara}
\frac{\beta_\perp}{\beta_\|} < 1 
\end{equation}
The smaller $\epsmax$ with respect to $m$ the more effective these constraints will be. (Notice that the functions are monotonous on the appropriate range.) 
 Moreover, the maximum value of $1-\cos\theta$ is the limiting factor for $\epsmax$ , and therefore these limits are less stringent if one considers creation of particles at smaller angles. Besides, $\sqrt{k\epsmax}$\footnote{The energy of one photon in the center of mass of two photons, one at energy $k$ and another at energy $\epsilon$, is $\sqrt{k\epsilon(1 - cos\omega)}$ where $\omega$ is the angle between the two photons 3-momenta} is the higher bound of the energy of the two photons in the center of mass frame, and given our condition $k \gg m$, $\epsmax$ close to $m$ leads to an energy way above threshold in Eq. (\ref{eq_threshold}), and therefore very unlikely to happen (section \ref{sec_two_photon}), although it depends on the angle of incidence of the weak photon on the strong one as well. For these reasons, we should consider that the higher limit for $\epsmax$ is a "smooth" one meaning that most photons of the weak distribution should actually not be close to $\epsmax$, even when $\epsmax$ is close to the limit $m/4$, except if one has a very peculiar photon distribution. This discussion a posteriori justifies condition \ref{mainapprox}. 
The proper vectors associated to the proper values in Eq. (\ref{propervalues}) are 
\begin{eqnarray}
\vec v_1&=&(0,0,1), \nonumber \\
\vec v_2&=&(0,1,v_{2z}), \nonumber \\
\vec v_3&=&(1,\frac{2 \beta_x \beta_y}{\beta_y^2-\beta_x^2)},v_{3z}), \nonumber
\end{eqnarray}
with 
\begin{eqnarray}
v_{2z}&=&\frac{2 \beta_y \beta_z}{\beta_z^2-\beta_y^2} << 1,\\
v_{3z}&=&\left[ 2 \beta_x \beta_z -4 \frac{\beta_x \beta_z \beta_y}{\beta_z^2-\beta_y^2}\right] \frac{1}{\beta_z^2-\beta_x^2} <<1.
\end{eqnarray}
The above components are negligible in virtue of Eqs. (\ref{domialpha}) and (\ref{dominpara}).
Therefore, any vector parallel to the $z$ axis has its image parallel to the $z$ axis, and any vector perpendicular to the $z$ axis has its image roughly perpendicular to the $z$ axis.

We can simplify the orthogonal proper values in Eq \eqref{propervalues}, which are now both equal to :
\begin{equation}
 \frac{\alpha^2}{A^2} 
\end{equation}
It can be shown that
\begin{eqnarray}
\label{eqalpahbetapara}
\alpha & < & \beta_\|, \\
\frac{\beta_\| - \alpha}{m} & =& \gdo{1}.
\end{eqnarray}
This implies that the parallel proper value (the third one in Eq. \eqref{propervalues}) is negative.
Because the parallel proper value is negative while the two orthogonal values are positive, the quadratic form in Eq.(\ref{eqsquareshell1}) describes a paraboloid of revolution.

%
%
%
%
%
%

\begin{figure}
	\begin{center}
		\def\svgwidth{0.5\textwidth}
	    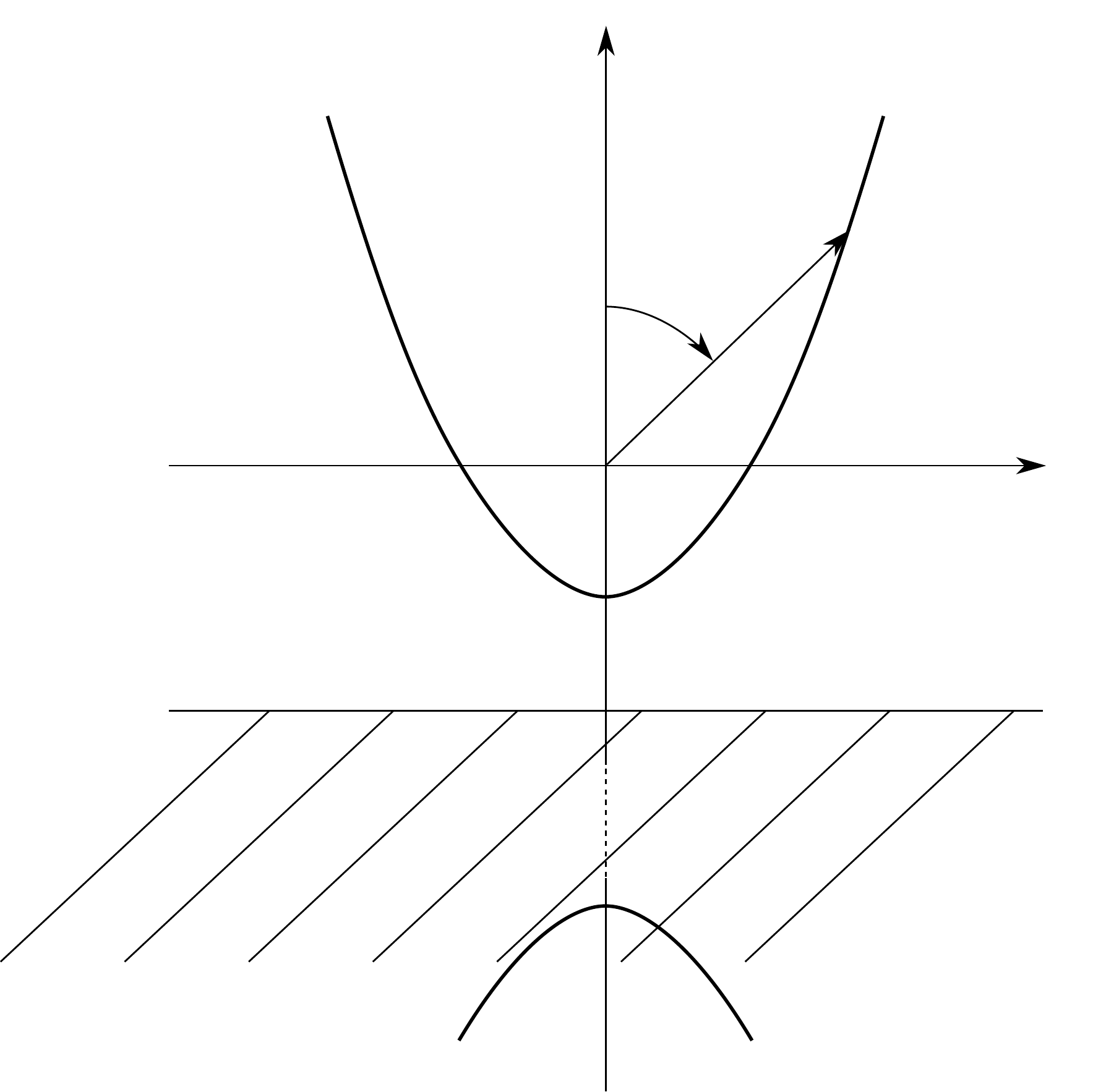
	 \end{center}
	 \caption{\label{figparaboloid} Representation of the mass shell \eqref{eqConstraintsLm}b) within approximation \eqref{mainapprox}.  }
\end{figure}

The above remarks and Eqs. (\ref{domialpha}) and (\ref{dominpara}) allow to simplify the quadratic form in Eq. (\ref{eqsquareshell1}).
We are left with 
%
%
\begin{equation}
\label{eqsquareshell3}
-\left(\frac{\beta_z^2}{A^2}  - \frac{\alpha^2}{A^2} \right)(z + z_0)^2 + \frac{\alpha^2}{A^2}(y^2 + z^2 ) = 1 - z_0^2\left(\frac{\beta_z^2}{A^2}  - \frac{\alpha^2}{A^2} \right),
\end{equation}
where
\begin{equation}
z_0 = \frac{A\beta_\|}{\beta_\|^2 - \alpha^2}.
\end{equation}
Dividing everything by $\left(\frac{\beta_z^2}{A^2}  - \frac{\alpha^2}{A^2} \right)$ we get our final, although not fully standard, form of Eq. (\ref{eqsquareshell1}):
\begin{equation}
\label{eqsquareshellfinal}
\frac{(z + z_0)^2}{z_0^2} - \frac{x^2 + y^2 }{L^2} = 1 - \delta^2,
\end{equation}
where the characteristic orthogonal radius $L$ and the displacement $\delta$ are
\begin{eqnarray}
L^2 & = & \frac{A^2}{\alpha^2}\frac{\beta_\|^2}{\beta_\|^2 - \alpha^2}, \\
\delta^2 & = & \frac{\beta_\|^2 - \alpha^2}{\beta_\|^2}.
\end{eqnarray}
Equation \ref{eqsquareshell3} describes a paraboloid of revolution of axis $\vz$, i.e. parallel to the strong photon 3-momentum $\vectk$. We notice that $L^2/z_0^2 = \gdo{1}$ meaning that the hyperboloid is very steep around the parallel axis. 
Besides, the small displacement $\delta$, $\delta^2 = \gdo{1}$, is responsible for a shift of the bottom of the paraboloid under the plane of zero parallel momemtum. This corresponds to reactions with head-on weak photons that are in general of smaller energies, as shown on figure \ref{figparaboloid}.

We must remeber that the inclusion of $L_-$ into the paraboloid defined in Eq. \eqref{eqsquareshellfinal} is derived from the condition in Eq. (\ref{eqConstraintsLm}b). It must be completed with the condition in Eq. (\ref{eqConstraintsLm}e) that reads
\begin{equation}
\label{eqsignconstrain}
  \vect{x} \cdot \vect{\beta}  + A \geq 0.
\end{equation}
From a geometrical point of view, this means that the relevant photons are those with momenta above the plane of normal vector $\vect{\beta}$ of equation $ \vect{x} \cdot \vect{\beta}  =- A $. Using relation \eqref{dominpara}, this approximates to the plane orthogonal to the parallel direction, $\vz$, at position :
\begin{equation}
 \zmin = - \frac{A}{\beta_\|} \simeq -2 \epsmin
\end{equation}
As a consequence, only the upper sheet of the hyperboloid defined in Eq. \eqref{eqsquareshellfinal} corresponds to the physical mass shell. Indeed this hyperboloid crosses the parallel axis $\vz$ at abscissa $z_\delta^\pm$ such that :
\begin{equation}
	z_\delta^\pm = -z_0\left(1 \pm \sqrt{1 - \delta^2}\right)
\end{equation}
This relation takes into account the fact that $z_0 \gg \abs{\zmin}$. 
Because $\delta <<1$,
\begin{equation}
	z_\delta = -z_0\left(1 - \sqrt{1 - \delta^2}\right) \simeq -\frac{1}{2}z_0\delta^2
\end{equation}
Further, one may show that $z_\delta = \min{\norm{\vx}}$ \footnote{Then one shows easily from \eqref{eqsquareshellfinal} that $x^2 + y^2$ has a minimum for $z_m = -z_0\frac{L/z_0}{1 + L^2/z_0^2} = -z_0 L/z_0 + \gdo{L^2/z_0^2}$ ($\gdo{L^2/z_0^2} = \gdo{1}$). Hence $z_m \simeq - \frac{A}{\alpha}\frac{\beta_\|}{\alpha}$ while $z_\delta = \frac{A}{\beta_\|}$. Using the fact that $\alpha \lesssim \beta_\|$ one gets that $z_m\lesssim z_\delta$, which means $z_m$ is slightly under the bottom of the hyperboloid. Since $x^2 + z^2$ can be easily shown to be a growing function of $z$, its smallest value can only be $z_\delta$.} which corresponds to the physical idea that the smallest weak photon that can produce a pair is the one that hits the strong photon head-on. How does it compare to $\epsmin$ determined in the previous section ? With the notations used in this section,
\begin{equation}
	\epsmin = \frac{A}{\norm{\vect{\beta}} + \alpha}.
\end{equation}
Using the approximation in Eq. (\eqref{dominpara}), $\norm{\vect{\beta}} = \beta_\| + \gdo{\beta_\perp/ \beta_\|}$, we get
\begin{equation}
	\epsmin = z_\delta +  \gdo{\beta_\perp/ \beta_\|}.
\end{equation}
This is consistent with the definition of $\epsmin$ in Eq. (\ref{eq_epsilonmin}) as the minimum energy allowed in $L_-$.

\subsection{Integration} \label{sec_integration}
We now compute the integral $W_k$ defined in Eq. \eqref{eqWoverV}. 
Within the frame of our approximations, Eq. \eqref{eqsquareshellfinal} shows that the probability of making a pair is symmetric around $\vectk$. This leads to a first angular integration of $\phi$ that yields a $\deuxpi$ factor. 
The differential element $\diftrois{\vect{k_w}} =\dif{x}\dif{y}\dif{z}$ is constrainted by Eq. \eqref{eqsquareshellfinal} that defines $L_-$.
%
Thus we write $z = z(x,y,p)$ through the constrain $L_-(p, \cos \theta,k)$ and
\begin{equation}
 \dif{z} = \abs{\pderiv{z}{p}} \dif{p}.
\end{equation}

We can now write \eqref{eqWapprox1} as follows
\begin{eqnarray} \nonumber
	W_{\vectk} = c &\deuxpi& \int_{C_1}^{C_2} \dif{\cos\theta} \int_{p_1}^{p_2} \dif{p} \times \\ 
	\label{eqWapprox2}
& &\int_{L_-(x,y)} \derivdeux{\sigma}{\Omega} \frac{K_s\cdot K_w}{K_s^0 K_w^0} f_w(\vect{k_w})\abs{\pderiv{z}{p}} \dif{x}\dif{y},
\end{eqnarray}
where $L_-(x,y)$ is the projection of $L_-$ onto the $(x,y)$ plane.
We need to expand the different quantities appearing in \eqref{eqWapprox2}. Let us start with an explicit projection of the upper hyperboloid on the plane $(\vx, \vy)$. This projection is a disc of radius
\begin{equation}
\label{eqR}
R = 2\epsmax\sqrt{\mu (1 - \mu)},
\end{equation}
where $\mu$ is defined in Eq. \eqref{eqmudef}.
For the change of variable $\theta \rightarrow \mu$, we need to switch the integration on $p$ with the integration on $\cos\theta$  in \eqref{eqWapprox2}, with 
\begin{equation}
\label{eqmujacobian}
	\dif{\cos\theta} =  2\frac{(k-p)}{p}\frac{\epsmax}{k}\dif{\mu}
\end{equation}
Moreover, the shape of the domain naturally suggests to use polar coordinates in the plane $(\vx, \vy)$, with radius $r = \sqrt{x^2 + y^2}$, angle $\phi_w$ and $\dif{x}\dif{y} = r\dif{r}\dif{\phi_w}$. 
The differential cross-section is
\begin{eqnarray}
\derivdeux{\sigma}{\Omega} = & -\frac{r_e^2}{4} \frac{p m^2}{k\epsmax^2\mu^2}  
\left[ \left( \frac{m^2}{4\epsmax p \mu} + \frac{m^2}{4\epsmax (k-p) \mu} \right)^2 - \right. 
\nonumber \\ \label{eqcrosssecapprox}
							 & \left.  
							 \frac{m^2}{4\epsmax p \mu} - \frac{m^2}{4\epsmax (k-p) \mu} - \unquart \frac{p}{k-p} - \unquart \frac{k-p}{p} \right].
\end{eqnarray}
The elementary current in Eq. \eqref{eqDefinitionJ} is
\begin{equation}
\label{eqcurrentapprox}
\frac{K_s\cdot K_w}{K_s^0 K_w^0} = 1 - \cos\xi + \gdo{\beta_\perp/\beta_\|},
\end{equation}
where
\begin{equation}
\label{eq:cosxi}
\cos\xi = 1 - 2\mu\frac{\epsmax}{\epsilon} + \gdo{1}.
\end{equation}
The expression of $\epsilon =  \sqrt{x^2 + y^2 + z^2} $ needs  as well to be developed as a function of $x$, $y$ and $p$, which implies to write a clear expression for $z(p)$. From Eq. \eqref{eqsquareshellfinal},
\begin{equation}
z = - z_0 + z_0 \sqrt{ 1 + \frac{x^2 + y^2}{L^2} - \delta^2  }.
\end{equation}
One can show that under approximations in Eq. \eqref{mainapprox}, $\frac{x^2 + y^2}{L^2} < 1/16$ and should be in practice much smaller. Therefore,
\begin{eqnarray}
\label{eqzapprox}
 z \simeq \frac{z_0}{2} \left( \frac{x^2 + y^2}{L^2} - \delta^2 \right)
\end{eqnarray}
This allows to make $\epsilon$ explicit,
\begin{equation}
\label{eqepsilon}
\epsilon = \frac{1}{4\mu\epsmax}\left(x^2 + y^2 + 4\mu^2 \epsmax^2 \right),
\end{equation}
as well as the derivative of $z$ :
\begin{equation}
\label{eqdzdp}
\pderiv{z}{p} = {\frac{k^2 }{2(k-p)p}\left(\frac{m^2}{kp} - \frac{2\epsmax\mu}{k}\right) }{\left(1 + \frac{r^2}{4\epsmax^2\mu^2}\right) }.
\end{equation}
We note 
\begin{eqnarray}
\nonumber 
\left(\pderiv{z}{p}\right)_{\text{left}} &=& {\frac{k^2 }{2(k-p)p}\left(\frac{m^2}{kp} - \frac{2\epsmax\mu}{k}\right) },
\\ \nonumber
\left(\pderiv{z}{p}\right)_{\text{right}} &=& {\left(1 + \frac{r^2}{4\epsmax^2\mu^2}\right) }.
\end{eqnarray}
We need the absolute value of $\partial{z}/\partial{p}$, and one can show from \eqref{eqdzdp} that it is always negative, so that we shall always take the opposite of Eq. \eqref{eqdzdp} and remove the absolute value in the following developments. 
%
The distribution function $f_w$  is the only element that depends on $\phi_w$. Moreover, the integration over the hyperboloid $L_-$ leads to get rid of $z$ through Eq. \eqref{eqzapprox}. For further developments, we explicitly keep track of the fact that $L_- = L_-(p,\cos\theta) = L_-(p,\mu) $
\begin{equation}
\label{eqaveragephiw}
	F_w(r, p, \mu) = \int_{\phi_w = 0}^\deuxpi f_w\left(r, \phi_w, z(r^2,\mu)\right) \dif{\phi_w}.
\end{equation}
We can separate the integration of \eqref{eqWapprox1} in several parts. The parts with a dependance on $r$ are to be found in the Jacobian $\abs{\partial{z}/\partial{p}}$ (see Eq. \eqref{eqdzdp}) of which we take only the rightmost factor, the current  in Eq. \eqref{eqcurrentapprox},  the distribution function, 
and the differential element $r\dif{r}$. Parts that depend only on $\mu$ or $p$ are the differential cross section in Eq. \eqref{eqcrosssecapprox}, the two first factors in the Jacobian $\abs{\pderiv{z}{p}}$ in Eq. \eqref{eqdzdp} , and the Jacobian in Eq. \eqref{eqmujacobian}. The dependance of the integrated distribution function $F_w$ is not known a priori. We obtain
\begin{eqnarray}
	\label{eqWapproxsplit}
	W_{\vectk} = c\deuxpi \int_{p_1}^{p_2} \dif{p} \int_{\mu_1}^{\mu_2} \dif{\mu} \pderiv{\cos\theta}{\mu} \derivdeux{\sigma}{\Omega}  \abs{\left.\pderiv{z}{p}\right|_{\text{left}}} \times 
	\\ \nonumber 
	\int_{r = 0}^{R}  \left.\pderiv{z}{p}\right|_{\text{right}} \frac{K_s\cdot K_w}{K_s^0 K_w^0} F_w(r, p, \mu)r \dif{r}.
\end{eqnarray}
The boundary conditions on the $p$ integral, $(p_i)_{i=1,2}$ must be understood as $p_i =  \max\left(p_i, k-p_i \right)$ in virtue of symmetry \eqref{eqWsymmetry}.
At lowest order, one can shows that the $r$ part of the integrand is merely equal to $ 2 F_w(r, p, \mu) r \dif{r}$ and that the $\mu$ part can be reduced after a partial fraction decomposition to 
\begin{equation}
 \sum_{i=1}^{4} \frac{a_i(p)}{\mu^i},
\end{equation}
where the dimensionless coefficients $a_i(p)$ are given by Eq. \eqref{eq_coefs_a_i}.
Then Eq.  \eqref{eqWapproxsplit} can be formally reduced to Eq. \eqref{eqWapproxsplitreduced}.

\section{Formula Compendium }
\label{apformulae}
The cosine of the angle $\theta$ between the strong photon $\vect{k}$ and the outgoing lepton $\vect{p}$ is parametrized below by 
\begin{equation}
\cos\theta = 1 - c
\end{equation}
and the following parametrization by $\mu$ can lead to significant simplifications 
\begin{equation}
  c = 2\frac{k-p}{p} \frac{\epsmax}{k}\mu - \frac{m^2}{2p^2} 
\end{equation}.

The following quantities are used are intermediates in the derivation of the hyperboloid of integration,

\begin{eqnarray}
A & = & K\cdot P \\
\vect{\beta} & = & \vect{k} - \vect{p} \\
\alpha & = & k - P^0 \\
\overline{\beta} & = & \left( \begin{array}{ccc}
\beta_x^2 & 0 & 0 \\
2\beta_x\beta_y & \beta_y^2 & 0 \\
2\beta_x\beta_z & 2\beta_y\beta_z & \beta_z^2
\end{array}\right)
\end{eqnarray}

and can be explicited to relevant order (see \eqref{eqbigon} ) as
\begin{equation}
\begin{array}{l}
\begin{array}{lcl}
A  & = &  m^2\left(\frac{1}{2}\frac{k}{p} + \frac{kp}{m^2}c\right)+ \gdo{1} = 2m^2 \frac{k-p}{m}\frac{\epsmax}{m}\mu \\
\alpha & = &    m\left( \frac{k}{m}-\frac{p}{m} - \frac{m}{2p} \right) + \gdo{2} \\
\end{array}\\
\begin{array}{rcl}
\abs{\beta_x}\sim\abs{\beta_y} \sim \sqrt{\beta_x^2 + \beta_y^2}  =  \beta_\perp & = & p\sqrt{2c} + \gdo{3} \\
\beta_z = \beta_\|  & = &  p\left(\frac{k}{p} - 1 + c\right)
\end{array}
\end{array}
\end{equation}.

The characteristics of the hyperboloid \eqref{eqsquareshellfinal}, are then related to the previous quantities by 
\begin{equation}
\begin{array}{ccccccc}
z_0 & = & \frac{A\beta_\|}{\beta_\|^2 - \alpha^2} & = & \frac{k}{2(k - p)} (k - p + pc) & = & \frac{k}{2} + \gdo{1} \\
L^2 & = & \frac{A^2}{\alpha^2}\frac{\beta_\|^2}{\beta_\|^2 - \alpha^2} & = & \frac{pk^2}{4(k - p)} \left(2c + \frac{m^2}{p^2} \right) & = & \mu k\epsmax \\
\delta^2 & = & \frac{\beta_\|^2 - \alpha^2}{\beta_\|^2} & = &\frac{p}{k - p + 2cp} \left(2c + \frac{m^2}{p^2} \right) & = & 4\mu\frac{\epsmax }{k} +\gdo{2}
\end{array}
\end{equation},
where $c = 1 - cos\theta$.

From this one finds the derivative of $z$  : 
\begin{equation}
\pderiv{z}{p} = \frac{k^2 }{2(k-p)p}\left(\frac{m^2}{kp} - \frac{2\epsmax\mu}{k}\right) \left(1 + \frac{r^2}{4\epsmax^2\mu^2}\right)
\end{equation}


\bsp	
\label{lastpage}
\end{document}